\documentclass[11pt]{article}
\usepackage{bm}
\usepackage{color}
\usepackage{graphicx}
\usepackage{xcolor}
\usepackage{multirow}
\usepackage{epstopdf}
\usepackage{geometry} 
\usepackage{slashed}
\usepackage{amsmath,amssymb,amsfonts}

\usepackage{fontawesome}
\usepackage[
    natbib=true,
    style=numeric,
    sorting=none
]{biblatex}
\usepackage{hyperref}
\DeclareFieldFormat{url}{\mkbibacro{URL}\addcolon\space\href{#1}{\faExternalLink}}

\usepackage{multirow}
\usepackage{tikz}
\usepackage{slashed}
\usepackage{geometry}
 \makeatletter
 \def\myline{\pgfutil@ifnextchar[{\my@line}{\my@line[]}}%
\def\my@line[#1](#2)(#3){%
\tikz[overlay] \draw[#1]  (#2)--(#3); 
}%
\usepackage{wrapfig}
\usepackage{bbold}

\usepackage{graphicx}
\usepackage[footnotesize]{caption2}
\usepackage{mathrsfs}
\usepackage{bbm}
\usepackage{braket}
\usepackage{cancel}
\usepackage{color}
\usepackage{mathtools}
\usepackage{enumitem}

\def\beq{\begin{equation}}
\def\eeq{\end{equation}}
\def\bea{\begin{eqnarray} }
\def\eea{ \end{eqnarray} } 

\newcommand{\eq}[1]{Eq.~\eqref{#1}}

\newcommand{\Figref}[1]{Fig.~\ref{#1}}
\newcommand{\Tabref}[1]{Tab.~\ref{#1}}

\usepackage{mathtools}

\newcommand{\eff}{\rm{eff}} 
\newcommand{\rsw}{\rm sw}
\newcommand{\turb}{\rm turb}

\allowdisplaybreaks

\newcommand{\SOT}{\text{SO(10)}}
\newcommand{\SUF}{\text{SU(5)}}
\newcommand{\SUT}{\text{SU(2)}}
\newcommand{\SUTH}{\text{SU(3)}}
\newcommand{\SUFR}{\text{SU(4)}}
\newcommand{\UO}{\text{U(1)}}
\newcommand{\SM}{\text{SM}}

\newcommand{\MG}{\:M_{\text{GUT}}}
\newcommand{\MBL}{\:M_{B-L}}
\newcommand{\MD}{\:M_{D}}
\newcommand{\MP}{\:M_{\text{P}}}
\newcommand{\MPS}{\:M_{\text{PS}}}
\newcommand{\MEW}{\:M_{\text{EW}}}
\newcommand{\MR}{\:M_{R}}

\usepackage{cleveref}

\crefformat{section}{\S#2#1#3} 
\crefformat{subsection}{\S#2#1#3}
\crefformat{subsubsection}{\S#2#1#3}

\newcommand{\suchthat}{\mathrel{\mathop\supset}\kern-4.0pt$-$\kern-2pt$-~$}

\makeatletter
\def\bstctlcite{\@ifnextchar[{\@bstctlcite}{\@bstctlcite[@auxout]}}
\def\@bstctlcite[#1]#2{\@bsphack
  \@for\@citeb:=#2\do{%
    \edef\@citeb{Expandafter\@firstofone\@citeb}%
    \if@filesw\immediate\write\csname #1\endcsname{\string\citation{\@citeb}}\fi}%
  \@esphack}
\makeatother

\newcommand{\xdownarrow}[1]{%
  {\left\downarrow\vbox to #1{}\right.\kern-\nulldelimiterspace}
}

\newcommand{\xuparrow}[1]{%
  {\left\uparrow\vbox to #1{}\right.\kern-\nulldelimiterspace}
}

\definecolor{olive}{rgb}{0.3, 0.4, .1}
\definecolor{fore}{RGB}{249,242,215}
\definecolor{back}{RGB}{51,51,51}
\definecolor{title}{RGB}{255,0,90}
\definecolor{dgreen}{rgb}{0.,0.6,0.}
\definecolor{gold}{rgb}{0.98,0.88,0.1}
\definecolor{orgold}{rgb}{0.92,0.7,0.1}
\definecolor{orange}{rgb}{0.9,0.4,0.1}
\definecolor{DarkBlue}{rgb}{0.2, 0.2, 0.6}

\begin{document}
\begin{flushright}
{\bf \small {CQUEST-2021-0676}}\\
{\bf \small {KIAS-P21064}}
\end{flushright}
\begin{center}
{\Large\bf 
Tracking Down the Route to the SM with Inflation and Gravitational Waves
}
\end{center}

\vspace{0.5cm}
\begin{center}{\large
{ Eung Jin Chun}$^1$,
and { L.~Velasco-Sevilla}$^{1,2,3}$\\
}
\end{center}

\begin{center}
{\em $^1$Korea Institute for Advanced Study, Seoul 02455, South Korea}\\[0.2cm]
{\em $^2$ Department of Physics, Sogang University, Seoul 121-742, South Korea}\\[0.2cm]
{\em $^3$ Department of Physics and Technology, University of Bergen,\\
PO Box 7803, 5020 Bergen, Norway}
\end{center}

\begin{abstract}
We explore supersymmetric SO(10) models predicting observable proton decay and  various topological defects which produce different shapes and strengths of gravitational wave backgrounds depending on the scales of intermediate symmetry breaking and inflation as well. We compare these to their non-supersymmetric counterparts.
By identifying the scales at which gravitational wave signals appear, we would be able to track down a particular breaking chain and discern if it has a supersymmetric origin or not. It would also be useful to observe gravitational waves from more than one source among all possible topological defects and first order phase transitions for a realistic breaking chain.  For these purposes, we work out specific examples in which the grand unification and relevant intermediate scales are calculable explicitly. It turns out that examples with gravitational waves from different sources are quite difficult to obtain, and the predicted gravitational wave profiles from domain walls and first order phase transitions obtained in some examples will require detectors in the kHz to MHz region.

\end{abstract}

\hypersetup{linkcolor=blue}
\tableofcontents

\section{Introduction}

The first ever direct detection of a gravitational wave (GW) signal  in 2015 \cite{TheLIGOScientific:2016agk} opened the era of gravitational wave astronomy and revived interests in cosmological GW signals. The study of GW signals from cosmological origin, either topological defects or phase order first transitions (FOPT), is a long standing subject, but only recently the current and future experiments have the potential to detect these kind of signals.
 During the next years,  KAGRA \cite{KAGRA:2020cvd} and LIGO-India will be integrated in the LIGO~\cite{Abbott:2016blz,Aasi:2014mqd,Thrane:2013oya,LIGOScientific:2019vic} and Virgo \cite{DiPace:2021hxc} interferometer network (LVC). Thanks to the improved sensitivity, a huge number of GW events will be detected. The detection could be so frequent that the associated signals would be too entangled to be identify individually. These events will then appear as a Stochastic Gravitational Wave Background (SGWB), a signal continuous in time, loud in a very broad frequency band, and coming from the whole sky dome. In fact, these are also the characteristics of GW signals created at the earliest moments of the universe. Such  signals, called SGWB of Cosmological Origin (SGWBoCO), exist but we do not know their strength and the frequencies at which we could detect them. It is then crucial to design SGWB searches suitable for scenarios where the SGWBoCO and SGWB of astrophysical origin (SGWBoAO) are both present at some level. 
This kind of searches will need to be implemented by all current and projected experiments, such as LISA\footnote{For updates on the LISA scientific prospects check \cite{Caprini:2019egz}.} 
TianQin~\cite{Luo:2015ght}, Taiji~\cite{Gong:2014mca}, LISA~\cite{Audley:2017drz}, Einstein Telescope \cite{Hild:2010id, Punturo:2010zz}, Cosmic Explorer~\cite{Evans:2016mbw}, BBO~\cite{Corbin:2005ny} and DECIGO~\cite{Yagi:2011wg}, as well as
 Pulsar Timing Array (PTA) experiments SKA~\cite{Janssen:2014dka}, EPTA~\cite{Desvignes:2016yex}, PPTA~\cite{Hobbs:2013aka}, IPTA~\cite{Verbiest:2016vem}.   In fact, in September 2020 the NANOGrav Collaboration
announced their PTA analysis~\cite{Arzoumanian:2020vkk} showing that we are likely on the verge of a SGWB detection.
  In particular, it has hinted at the detection of a cosmic string in the range $G\mu \in[  2\times 10^{-11}, 3 \times 10^{-10}]$ \cite{NANOGRAV:2018hou}.
A firm SGWB observation may require only some extra years of measurements, and it is encouraging to have the recent positive development of the PPTA analysis \cite{Goncharov:2021oub} and the collaborative effort of IPTA \cite{Antoniadis:2022pcn}.
If confirmed, the NANOGrav SGWB  signal in the range $G\mu \in[  2\times 10^{-11}, 3 \times 10^{-10}]$ would be compatible with the SGWBoCO produced by (Nambu Goto) cosmic strings coming from a phase transition that occurred between $10^{-22}$ and $10^{-18}$ seconds after the Big Bang. 

On the theoretical side, model building within the context of Grand Unified Theories (GUT) needs to provide a compelling science case for scenarios where  the cosmological signal can be clearly identified.  In this respect, there has been a large number of recent studies reviving the study of GW coming from topological defects \footnote{Relevant discussions previous to the GW detection include \cite{Jeannerot:1995yn,Jeannerot:2000sv,Jeannerot:2003qv,Jeannerot:2005ah,Sakellariadou:2005wy,Sakellariadou:2007bv}. }.
These include  non-supersymmetric GUT theories in connection to proton decay and unification
\cite{Chakrabortty:2017mgi,Chakrabortty:2019fov, King:2021gmj} and different interconnections to $B-L$ \cite{Buchmuller:2019gfy}, neutrino masses and leptogenesis \cite{Dror:2019syi}, as well as in connection to addressing the NANOGrav hint for cosmic strings \cite{Chakrabortty:2020otp, Lazarides:2021uxv} and combined effects with FOPT and GW  \cite{Zhou:2020ils}.

In this paper, we address the question if different breaking chains of the supersymmetric SO(10) GUT predicting a certain combination of topological defects and first order phase transition (FOPT) can be identified through the detection of SGWB.  This task of course can be done taking into account low energy phenomenological constraints (proton decay, masses, flavor effects, etc.) and assumptions about how inflation should take place.  

The classification of breaking routes of SO(10) down to Standard Model (SM), or Minimal Supersymmetric SM (MSSM), producing topological defects, started  long time ago \cite{Jeannerot:2003qv}.
As it is well-known, non-supersymmetric models can yield gauge coupling unification with different intermediate scales \cite{Rajpoot:1980xy} and therefore it is appealing to study the GW panorama in this context. 
Making a catalogue of surviving breaking chains coming from SO(10) models as in  \cite{Chakrabortty:2019fov,King:2021gmj} {\footnote{Even in these cases, a certain number of assumptions are needed. For example, to consider or not threshold corrections to achieve gauge coupling unification and acceptable proton decay rates, and also to consider or not multiple  parameters controlling  topological defects \cite{Martin:1996ea,Vilenkin:2000jqa}.} is considerably more challenging in supersymmetry due to the fact that not only the number of constraints and parameters is considerably bigger but also to the assumptions for the way supersymmetry is broken and the way it is  UV completed. For these reasons we present in this work only a few typical examples, rather than a comprehensive catalogue.

Supersymmetric models can be compared with their non-supersymmetric counterparts because they differ in their breaking scales and proton decay channels and this can be contrasted in the future with both GW and proton decay experiments.
We provide a study
for this comparison by considering examples based on the SO(10) breaking routes containing the $\SUF$ and  $\SUTH_C \times \SUT_L \times \SUT_R \times \UO_{B-L}$ sub-groups. These models are representatives of different subgroups and low energy phenomenology and therefore can shed light on discerning routes down to the SM.  We envision plots for frequency vs. density showing all possible combination of topological defects and FOPT for a particular breaking chain. Unfortunately, this task is quite model-dependent and will require future developments of simulations, in particular, for hybrid topological defects since the relative tensions of the topological defects is crucial to the evolution of the networks. Nevertheless, it is worth to establish possible breaking routes where hybrid topological defects can leave a distinct GW imprint. Recently, these observations have been also revived in \cite{Dunsky:2021tih} with a similar motivation to ours.

The paper is organized as follows. In \cref{sec:GWGUTS} we put into context the appearance of topological defects according to the different breaking routes down to the SM.
In \cref{sec:phenogral} we mention how do we obtain the scale of breaking and how do we compute the proton decay ratios.  In \cref{sec:bcPS} we analyze our model examples and compare to the non-supersymmetric counterparts.  In \cref{sec:conclusions} we conclude. For the sake of completeness, we provide the information that we use to generate the GW signals in the appendices.

\section{GUT breaking and topological defects \label{sec:GWGUTS}}

In this section, we make a brief summary of topological defects appearing in a GUT breaking chain which we use and refer to. Then, we introduce typical SO(10) GUT breaking patterns for which we study various phenomenological features and GW signals from relevant topological defects and possible phase transitions.  

\subsection{Theory of topological defects in breaking chains}

A schematic breaking chain of a GUT model down to the SM can be depicted as
\bea
\label{eq:tpdchain}
G_{\text{GUT}}\ \xrightarrow[]{p_1,p_2,\hdots} \ H_1  \xrightarrow[]{q_1, q_2, \hdots} \ H_2\ \hdots \ \xrightarrow[]{r_1, r_2, \hdots} \ H_n \ \xrightarrow[]{s_1, s_2, \hdots}  \ G_{\rm{SM}},
\eea
where the letters $p_i$, $q_i$ and $r_i$ represent topological defects.  The conditions for the formation, evolution and stability of topological defects has been thoroughly studied \footnote{See for example \cite{Preskill:1992ck,Vilenkin:2000jqa} for comprehensive reviews. Of course improvements for simulations and further understanding are currently undergoing.}. 
The $k-$homotopy group, $\pi_k(G/H)$,  of the vacuum manifold $\mathcal{M}=G/H$  determines the appearance of topological defects since $\pi_k(G/H)$ classifies the distinct topological spaces of $G/H$ that appear after the breaking of $G\rightarrow H$. They correspond to $k=3$ for textures, $k=2$ for monopoles, $k=1$ for cosmic strings (CS) and $k=0$ for domain walls (DW), although formally $\pi_0$ is not a group: it represents merely the number of connected components of the manifold. If the homotopy groups  are trivial, that is, $\pi_k(G/H)=I$, then there are no  associated defects. The general features of topological defects from the breaking sequence of \eq{eq:tpdchain} can be summarized as follows.

\begin{enumerate}
    \item The topological defects will be stable up to the $H_n$ group if the k-homotopy group of the manifold $G_{\rm{GUT}}/H_n$ is not trivial, that is if $\pi_k\left(G_{\rm{GUT}}/H_n\right)\neq I$;
    and  unstable up to the $H_n$ group if the k-homotopy group of the manifold $G_{\rm{GUT}}/H_n$ is trivial, that is  if $\pi_k\left(G_{\rm{GUT}}/H_n\right)= I$.
    The same applies for the manifold $G_{\rm{GUT}}/\left(SU(3)_C\times U_Y(1)\right)$ which will determine stable or unstable topological defects all the way down to the EW scale. 
    \item There can appear metastable defects that decay quantum mechanically  with a decay rate  $\propto e^{-{\mathcal{A}}/\hbar}$, where 
    $\mathcal{A}$  is the ``tunneling action" of the particular defect     \cite{Preskill:1992ck}.
     For the strings decaying via a pair of monopole/anti-monopole the decay rate, 
     which is specifically the tunneling rate per unit string length, $\ell$, is given by
     \bea
     \label{DR:CS}
     \frac{dP}{d\ell}=\Gamma= \frac{\mu}{2\pi} e^{-\pi\, k}, \quad k=\frac{m^2}{\mu},\, 
     \eea
     where $\mu$ is the CS tension and $m$ is the scale at which the monopoles are begin created. This means that the creation of strings should not take place much below the formation of monopoles as only decays for values of $k=O(1)$ render a non negligible value of $\Gamma$, since the exponential in \eq{DR:CS} decays rapidly.  For values much greater than 1, the decaying cosmic strings become indistinguishable from stable strings.
     In certain cases, the probability per unit area of a domain wall decaying by the nucleation of strings, goes like \cite{Kibble:1982dd,Preskill:1992ck}
      \bea
      \label{DR:DWnuc}
      \frac{dP}{dA} \sim \sigma\, e^{- 16\pi /3 \, \mu^3/\sigma^2},
      \eea
       where $P$ is the probability of nucleation and $A$ is the unit area, $\sigma$ and $\mu$ are respectively the tension of the domain wall and the cosmic string.
    Hence, depending on the tunneling process or the decay time, the metastable defects could appear as (almost) stable or disappear quickly and hence leave different GW background imprints.
    \item Hybrid topological defects appear when a defect associated with the homotopy $\pi_k(H_i/H_j)$ is produced at a breaking stage and at a subsequent breaking stage a defect with a homotopy $\pi_{k-1}(H_j/H_\ell)$ is generated. Then, the topological defects associated to $\pi_k(H_i/H_j)$ interact with those associated with $\pi_{k-1}(H_j/H_\ell)$ \cite{Kibble:1982ae}. Examples of these phenomena include the well known fact that monopoles could be the seeds of monopole-antimonopole decay of a string via the Schwinger mechanism and also the fact that strings could become the boundaries of the DW produced at a later stage  \cite{Kibble:1982ae,Kibble:1982dd,Everett:1982nm}.     The interaction and evolution of these defects depend on the tensions and sizes of the interacting defects and on when inflation and reheating take place (see for example \cite{Vilenkin:2000jqa} for a comprehensive review).
    \item If monopoles and cosmic strings are produced at the same breaking step or without any sizable gap, they are unstable \cite{Vilenkin:1982hm}, but could leave observable GW spectra \cite{Martin:1996cp,Martin:1996ea}.
    \item If domain walls appear together with cosmic stings after inflation,  they become unstable due to the attachment of strings and thus leave GW signals without causing a cosmological problem. Furthermore. a discrete symmetry associated with a domain wall could be lifted by Planck mass suppressed terms \cite{Rai:1992xw} and thus the domain wall decays to leave an observable GW signature while its cosmological evolution does not conflict with the standard cosmology\cite{Mishra:2009mk}.
    \end{enumerate}

Due to the recent development in GW detection, there have been extensive studies of the signatures from the GUT breaking.
Distinct GW signals from cosmic strings produced in association with/without  monopoles, or around/away from the end of inflation could be detected to provide
an important clue of both inflation \cite{Sarangi:2002yt,Copeland:2003bj} and GUT theories.
A typical  $\UO$ sub-group of SO(10) GUT is  $\UO_{B-L}$ whose breaking at different scales would be tracked by distinguishing the resulting SGWB \cite{Croon:2018kqn,Huang:2020bbe}.  On the other hand,
 the formation of domain walls and the GW production could be a consequence of the spontaneous breaking of discrete R symmetries \cite{Takahashi:2008mu,Dine:2010eb}.
There have been some arguments against the possibility of finding R symmetries \cite{Fallbacher:2011xg} in 4D GUT theories.  Lastly, the breaking of supersymmetry itself it can lead to FOPT \cite{Vaknin:2014fxa} that can generate GW \cite{Craig:2020jfv}. 

This makes the study of GW in the context of GUT a fascinating arena that can probe some of the fundamental aspects and predictions of GUTs. As mentioned before, the study of topological defect formation, evolution and production of GW signals has a long history, however  more elaborated simulations and treatment of signals are taking place recently given the opportunity for current and future experiments.~\footnote{See for example \cite{Buchmuller:2021mbb} for some recent considerations for the evolution of metastable strings.}

\subsection{Typical examples of SO(10) breaking}

Here  we present  possible breaking chains having more than one source of GW signals  and that we will analyze in \cref{sec:bcPS}. As it is well-known, uncertainties in determining the parameters involved in the processes of GW production are yet to be greatly reduced. Nevertheless, we can still make qualitative analyses revealing important features.
Let us now depict the distinct breaking chains for which we determine
the intermediate and unification scales and estimate proton decay ratios:
\bea
\SOT\!\! & & 
\xrightarrow[126~(\mbox{\scriptsize{2}}) ]{\MG} \SUF\times \UO_V  \xrightarrow[45~(\mbox{\scriptsize{2}})]{\MR} \
\SUTH_C \times \SUT_L \times \UO_V \times \UO_Z, 
\xrightarrow[45~(\mbox{\scriptsize{1}})]{\MBL} G_{\SM} \times \mathbb{Z}_2 \nonumber\\
&&\hspace*{8cm} \xrightarrow[10]{\MEW} \ \SUTH_C \times \UO_Y \times \mathbb{Z}_2,
\label{bk:so10_126}\\
\SOT\!\! & & \xrightarrow[45~(\mbox{\scriptsize{2}}) ]{\MG}
\small{\SUTH_C \times \SUT_L \times \SUT_R\times \UO_{B-L}}\xrightarrow[126~(\mbox{\scriptsize{1,pt}})]{\MBL}
\, G_{\SM}\times \mathbb{Z}_2 \nonumber\\
&&\hspace*{8.0cm}\xrightarrow[10]{\MEW} \ \SUTH_C \times \UO_Y \times \mathbb{Z}_2, \label{bk:case210URBL}\\
\SOT\!\! & & \xrightarrow[210~(\mbox{\scriptsize{2,1}})]{\MG}
\small{\SUTH_C \times \SUT_L \times \SUT_R\times \UO_{B-L} \times D}\xrightarrow[45~(\mbox{\scriptsize{0}})]{\MD} \nonumber\\
& &  \small{\SUTH_C \times \SUT_L \times \SUT_R\times \UO_{B-L} }
\xrightarrow[126~(\mbox{\scriptsize{1}})]{\MBL} G_{\SM}\times \mathbb{Z}_2 \nonumber\\
&&\hspace*{8.0cm}\xrightarrow[10]{\MEW} \ \SUTH_C \times \UO_Y \times \mathbb{Z}_2. \label{bk:case45URBL}
\eea
The numbers (2,1,0) in parentheses denote respectively again topological defects  (monopole, string, domain wall), and (pt) means phase transition.  The subscripts of the subgroups in Eqs.~(\ref{bk:so10_126})-(\ref{bk:case45URBL}) are as follows: C for color, L for Left, R for right,  $V = 4\, I_R^3-3(B-L)$ and $Z= -I_R^3 + 1/2(B-L)$. Where $I_R^3$ is the third component of the right-handed isospin and $B - L$ the baryon minus the lepton number. 
The notation $D$ represent the D parity also known as $Z_2^C$. 
Note that $\mathbb{Z}_2$, which is used in supersymmetry as R-parity, is never broken.
 Here $\MG$, $\MR$ and $\MBL$ refer respectively to the GUT scale, the scale where $\SUT_R$ is broken and the scale where $B-L$ is broken. 
 The forthcoming results of JUNO\cite{JUNO:2015zny}, DUNE \cite{DUNE:2020ypp} and Hyper-Kamiokande \cite{Hyper-Kamiokande:2018ofw} experiments will be able to tell us which models are going to be ruled out and therefore this will be able to exclude the models as sources of GW at a particular scale.  
 
One may consider the breaking route of \eq{bk:case210URBL} or \eq{bk:case45URBL} preceded by the Pati-Salam (PS) group $\SUFR_C\times \SUT_L \times \SUT_R$, namely adding a PS breaking scale $\MPS$ between $\MG$ and $\MBL$ or $\MD$. However, we find no attainable solutions in the minimal models with $\MG$ around ($10^{15},10^{16}$) GeV unless $\MG$ and $\MPS$ become the same  due to the restricted beta function coefficients of the PS group in the supersymmetric version of the model. More details are provided in Appendix B.

It is indeed interesting to look for breaking chains where combined sources of gravitational waves could appear. It turns out, however, that this does not occur generically.  For supersymmetric models with one intermediate step of breaking it is difficult to separate the intermediate scale too much from the unification scale without incurring proton decay problems.  With more than one intermediate scale,  an intermediate scale could be lowered, e.g., between $10^{10}$ GeV and  $10^{13}$ GeV.  Then a further splitting of scales can be achieved and combined effects can take place. This is the case for the example in (\ref{bk:case45URBL}).
For non-supersymmetric models, the separation of the GUT scale $\MG$ and any other intermediate scales, $M_{\text{I}}$, $M_{\text{II}}$, is more feasible basically because the number of particles in the theory, which change the coefficients in the beta functions, are much larger than in their supersymmetric counterparts and therefore one can allow more running of the gauge couplings between the intermediate scales. However, it is precisely this hierarchy, i.e. $\MG \gg M_{\text{I}}$ (or any other intermediate scale $M_{\text{II}}$), that erases potential interactions of combined effects. 
Specifically, the decay probabilities of Eqs.~ (\ref{DR:CS}) and
(\ref{DR:DWnuc}) decrease very rapidly because the arguments in the exponential go like $-1\times $ ratios of powers of the masses involved in the breaking chains. For example, for a decaying string at a scale  $\sim M_{\text{I}}$, the factor $k$ in \eq{DR:CS} goes like the ratio of $\MG^2/M^2_{\text{I}}$,  assuming that monopoles are created at $\MG$. We also note that the example in (\cref{bk:case210URBL}) allows the signatures both from cosmic strings and from FOPT. However, the signals from domain walls and FOPT turn out to be away from the current sensitivity as it will be shown in \cref{sec:bcPS}.

\section{Phenomenological considerations \label{sec:phenogral}}

The minimum set up for studying GUT, and its different breaking chains, calls for computation of unification scales and proton decay depending on the choices of matter content and intermediate scales.  Detailed aspects for this are well studied in the context of non-supersymmetric models. When supersymmetry is considered, one needs to make more assumptions about the spectra and boundary conditions. We will take no-scale supergravity boundary conditions following the treatment in \cite{Ellis:2019fwf}.
 
\paragraph{Gauge coupling unification}
Except for the cases studied previously in the literature, we compute the beta functions following \cite{Martin:1993zk} and express them in the usual form
$\mu \, dg_a/d\mu= b_a^{(1)}/16/\pi^2 \, g_a^3$
$+ g_a^3/(16\pi^2)^2$  $\, \left[\sum_{b=1}^3\, b_{ab}^{(2)}\, g_b^2 - c_a y^2_t \right]$, where we keep the top Yukawa coupling, $y_t$, since due to its relative size in comparison to the gauge couplings it cannot be neglected.

 We assume the ``survival hypothesis'' implying that only the SM particles contribute to the running below the  intermediate symmetry breaking scale and all other particles have masses around either $\MG$ or the intermediate scales. 

\paragraph{Proton decay} For non-supersymmetric theories the most sensitive channel is the dimension-6
operator induced decay $p\rightarrow \pi^0  \, e^+$ mediated by gauge fields. 
We estimate the proton decay for this channel using \cite{Mambrini:2015vna}

\bea
\label{eq:pdePi0}
\Gamma(p\rightarrow \pi^0  \, e^+)&=&
\frac{m_p}{32\pi}\left(1-\frac{m^2_{\pi^0}}{m^2_p} \right)^2\, 
\left[ |\mathcal{A}_L(p\rightarrow \, \pi^0 \, e^+ )|^2 + |\mathcal{A}_R(p\rightarrow  \, \pi^0 \, e^+)|^2
\right],
\eea
where $m_p$ and $m_{\pi^0}$ are the proton and the neutron pion, $\pi^0$,  masses respectively. The amplitude at the weak scale is computed from
\bea
\label{eq:Amplitudes_NS_PD}
{\mathcal{A}_L}(p\rightarrow \pi^0  \, e^+) & =& C_{RL}((ud)_R u_L)(\mu=2\, \rm GeV) 
\langle {\it{ 
\pi}}^0{\it{| (ud)_R u_R|p}}
\rangle, \nonumber\\
{\mathcal{A}_R}(p\rightarrow \pi^0  \, e^+)  &=& 2 C_{LR}((ud)_L u_R)(\mu=2\, \rm GeV) 
\langle {\it{
\pi}}^0{\it{| (ud)_R u_R|p}}
\rangle,
\eea
where the Wilson Coefficients  $C_{RL}((ud)_R u_L)$ and $C_{LR}((ud)_L u_R)$ corresponds to $C_1$ and $C_2$ of
\cite{Mambrini:2015vna}, respectively. For numerical values we use the inputs given in \Tabref{tbl:tablMzvals}.
The most constraining channel for the supersymmetric theories is $p\rightarrow K^+ \bar{\nu}$, mediated by dimension-5 operators.  We use \cite{Ellis:2019fwf}
\bea
\label{eq:GampKnu}
\Gamma(p\rightarrow K^+ \bar{\nu}) = 
\frac{m_p}{32\pi} \left( 1-\frac{m^2_K}{m^2_p}
\right)^2 | \mathcal{A}(p\rightarrow K^+ \bar{\nu_i})|^2,
\eea
where $m_K$ is the mass of the kaon and the decay amplitude is given in terms of the Wilson Coefficients effective at the hadronic scale (2 GeV)
\bea
\label{eq:ApKnuWC}
\mathcal{A}(p\rightarrow K^+ \bar{\nu_i}) &=&
C_{RL}(usd\nu_i)\langle K^{+}| (us)_R\, d_L   | p \rangle + 
C_{RL}(uds\nu_i)\langle K^{+}| (ud)_R\, s_L   | p \rangle  \nonumber\\
&+ &
C_{LL}(usd\nu_i)\langle K^{+}| (us)_L\, d_L   | p \rangle + 
C_{LL}(uds\nu_i)\langle K^{+}| (ud)_L\, s_L   | p \rangle.
\eea
It is useful to remember that all of the Wilson Coefficients in \eq{eq:ApKnuWC} are suppressed by the color-triplet Higgs mass and the supersymmetry breaking scale.
The hadronic matrix elements are given in \Tabref{tbl:tablMzvals} and the evolution of the Wilson coefficients from $\MG$ is given in \cite{Ellis:2019fwf}.
\begin{table}
\small
\begin{center}
\begin{tabular}{|l|c|c|c|}
\hline
\multicolumn{4}{|c|}{Experimental/Lattice values used}\\
\hline
Parameter & Current Value & Projected Bound & $3\, \sigma$ Discovery\\
\hline
$M_Z(M_Z)$ &  91.1876 (21) GeV & &\\
$\alpha_{S}(M_Z)$ & 0.1179(10) & &\\
$\alpha_{\rm{e.m.}}(M_Z)$ & 1/127.9 & & \\
$\sin^2\theta_W(M_Z)$ & 0.23121(4)& &\\
$m_t(m_t)$ & $172.5 \pm 0.7$ & &\\
$\langle \pi^0 | (ud)_R\, u_L| p\rangle$ & $-0.131(4)(13)$ \cite{Aoki:2017puj}  & &\\ 
$\langle \pi^0 | (ud)_L\, u_L| p\rangle$ & $0.134(5)(16)$ \cite{Aoki:2017puj} & &\\ 
$\tau (p\rightarrow  \, \pi^0\,  e^+)$ &  $ > 1.6 \times 10^{34}$ yrs \cite{Aoki:2017puj} 
\cite{Super-Kamiokande:2016exg} & $7.8 \times 10^{34}$ yrs $90\%$ C.L {\small{[HK\cite{Hyper-Kamiokande:2018ofw}]}}&  $6.3 \times 10^{34}$ yrs {\small{[HK\cite{Hyper-Kamiokande:2018ofw}]}}\\
 & $95\%$ C.L  & & \\
$\tau (p\rightarrow \, K^+ \, \bar\nu)$ &  $> 6.6 \times 10^{33} \quad $yrs 
& $5\times 10^{34}$ yrs {\small{[DUNE \cite{DUNE:2020ypp}]}} &\\
 & $95\%$ C.L  & $3.2 \times 10^{34}$ yrs $90\%$ {\small{C.L. [HK\cite{Hyper-Kamiokande:2018ofw}]\hspace*{-0.3cm} }}& $2 \times 10^{34}$ yrs  {\small{[HK\cite{Hyper-Kamiokande:2018ofw}]}} \hspace*{-0.3cm}\\
\hline 
\end{tabular}
\end{center}
\caption{Experimental/lattice values used to obtain values of gauge coupling unification and proton decay rates. \label{tbl:tablMzvals}}
\end{table}
{The forthcoming experiments DUNE \cite{DUNE:2020ypp}, JUNO \cite{JUNO:2015zny} and Hyper-Kamiokande \cite{Hyper-Kamiokande:2018ofw} are expected to improve the current sensitivity
 by an order of magnitude  (see \Tabref{tbl:tablMzvals}).}

\section{Results for Different Breaking Chains\label{sec:bcPS}}

\subsection{$\mathbf{SU(5)}$ routes\label{sec:bcsu5}}

 Non-supersymmetric models via SU(5) routes are quite restricted by the proton decay bound in the channel $p\rightarrow \pi^0 \, e^+$ since the Wilson coefficients are suppressed only by the masses of heavy gauge fields: $C_i\propto 1/M^2_X$ (for a review see for example \cite{Langacker:1980js}.) 
 Thus, the discussion below is  specific to supersymmetric SO(10).
 The following routes are candidates for the appearance of combined effects:
\bea
&& \SOT \, \rightarrow \, \SUF \times \mathbb{Z}_2 \, \rightarrow \,  G_{\SM}  \times  \mathbb{Z}_2, \label{eq:su5r1}\\
&& \SOT \, \rightarrow \, \SUF_F \times \UO_V\, \rightarrow \, G_{\SM}  \times  \mathbb{Z}_2, \label{eq:su5r2}\\
&& \SOT \, \rightarrow \,  \SUF \times \UO_V \, \rightarrow \,  \SUTH_C \times \SUT_L \times \UO_Z \times \UO_V  \, \rightarrow \,  G_{\SM} \times  \mathbb{Z}_2.\label{eq:su5r3} 
\eea
The first one, \eq{eq:su5r1}, produces  cosmic strings at the first step of breaking, and monopoles at the second stage of breaking. As  monopoles need to be inflated away, the only GW signal would be due to a possible breaking of R-parity ($\mathbb{Z}_2$) or FOPT. 
The second possibility, \eq{eq:su5r2}, produces monopoles at the first stage of breaking and  cosmic strings at the 
second stage of breaking.  However, the proton decay bound requires both breaking scales to be at the same scale close to $10^{16}$ GeV. Thus, inflation needs to take place after that, leaving no imprint from GW, other than the possible $\mathbb{Z}_2$ breaking.  Note however that the situation can be different in the flipped models.  The breaking of $\SUF_F \times \UO_V$ (and thus one $\UO$ factor) can occur at a much lower scale than $10^{16}$ GeV, if we introduce a pair of vector-like particles \cite{Huang:2006nu}.
For the third route, \eq{eq:su5r3}, monopoles will be produced in the first two breaking steps, while cosmic strings in the last one. Here unification also requires all the three scales to be close to $10^{16}$ GeV.
However, just as in the case of of the flipped SU(5) pair of vector-like fields could be introduced to split the scales \cite{Huang:2006nu}. 

In supersymmetric models, the dimension-5 operator in the channel $p\rightarrow K^+ \bar{\nu}$  is mediated by the exchange of colour-triplet Higgs and suppressed by the scale of supersymmetric particles, $M_S$, so effectively the operators mediating proton decay are proportional to  $\propto 1/(\MG \, M_S)$. This means that we can push up the supersymmetric scale to aid overcoming the present bounds \cite{Hisano:2013exa,Ellis:2019fwf}.  To have an idea we can write, \cite{Langacker:1980js},
\bea
\label{eq:Tp_estimate_gen}
\tau  \sim 4 \times 10^{35} \times \sin^4( 2\beta) \times \left( \frac{0.1}{\overline{A}_R}\right)^2 \,
\left( \frac{M_S}{100 \, \rm{TeV}}\right)^2 \left( \frac{M_{HC}}{10^{16} \, \rm{GeV}} \right)^2\, \rm{yrs.},
\eea
where $M_S$ is the supersymmetry breaking scale, $M_{HC}$ is the colour-triplet Higgs mass, $\overline{A}_R$ is a hadronic parameter and $\beta$ is the usual angle determined by the ratio of the vacuum expectation values of the two Higgs doublets.
For $M_S=10$ TeV, we get a suppression of $O(10^{-2})$ in \eq{eq:Tp_estimate_gen} making the proton decay just compatible with the current bound $\tau(p\rightarrow K^+ \overline{\nu} )  > 6.6 \times 10^{33}$ yrs and thus accessible by the upcoming experiments \cite{JUNO:2015zny,DUNE:2020ypp,Hyper-Kamiokande:2018ofw}.  
However, it is nontrivial and it is quite model dependent to achieve the measured Higgs mass $m_h=125 $ GeV.
This puts additional constraints on the model.

 When $\UO_V$ and $\UO_Z$ are broken, there can appear $B-L$ proton decay modes through, for example, the  operator  $d^c d^c u^c L H_u$ in supersymmetry \cite{Babu:2012vb}.  However, the resulting proton decay rate depends on the flavour structure of the model and can be suppressed sufficiently. For our analysis, we adopt the results of Fig.~3 of \cite{Ellis:2019fwf} showing the regions where the proton decay rate is compatible with the bound of the channel $p\rightarrow K^+ \bar{\nu}$. These regions correspond generally to the mass range of $M_{1/2} \geq 5 $ TeV and $M_0 \geq 7 $ TeV. Taking  $(M_{1/2},M_0,\tan\beta,\mu)=$ $(5.5\, \rm{TeV}, 7.0\, \rm{TeV}, 5, >0)$, we have the GUT scale of 
 \bea
 \label{eq:scalessusyEx2}
 \MG &=& 8.2 \times 10^{15}\,  \text{GeV},
 \eea
 and a proton decay rate of
 \bea
 \tau(p\rightarrow K^+ \bar{\nu})= 8.6 \times 10^{33}\  \rm{yrs}.
\eea
This would requires to break all the chains of Eqs.~(\ref{eq:su5r1}-\ref{eq:su5r3}) directly to $G_{SM}\times \mathbb{Z}_2$ at the GUT scale. 

The intermediate scales can be separated away from the GUT scale by altering the unification scale and introducing an effective intermediate scale with the addition of multiplets which affect a bit the running but do not have a great impact in the proton decay rate. This would allow detectable hybrid topological defects appearing in the SU(5) route. 
As an example, we take the chain  \eq{eq:su5r3} and consider the following two cases:
\bea
\MG &=&\MR \sim 10^{16} \, {\text{GeV}},  \  \MBL > 10^{14}\, {\text{GeV}},\label{eq:hierarchy1SU5} \nonumber\\
\MG&=&10^{16} \, {\text{GeV}}, \, \MR > 10^{14}\, {\text{GeV}}, \quad \MBL \sim  10^{14} \, {\text{GeV}}.\label{eq:hierarchy2SU5}
\eea
Depending on when the inflation takes place, we may observe different gravitational wave signals. If the reheat temperature $T_{RH}$ is higher than $\MBL$, there appear the usual signal of undiluted stable comic strings. In the opposite case $T_{RH}<\MBL$, all the defects are diluted away, but sizable string regrowth may occur depending on the conditions.  They are basically determined by choosing the parameters that satisfy Eq.~(7) of \cite{Cui:2019kkd}.  A brief summary about this is provided in the Appendix (\ref{app:ProdGWCS}). Basically we assume a number of long strings with the parameter choice of $\tilde{z}=10^{4}$ which is used in the plot \Figref{fig:su5EA}. 
Another interesting phenomenon is the appearance of observable signals of decaying cosmic strings attached with monopole-antimonopole pairs. 
The second breaking route of \eq{eq:hierarchy2SU5} can realize this possibility as  $\MR$ is slightly above than $\MBL$ and the inflation can take place in between, $\MR>T_{RH}>\MBL$.
We follow \cite{Leblond:2009fq} in order to get the profile of the decaying cosmic string.
All of these features are illustrated  in \Figref{fig:su5EA}.
Our breaking chain also allows the possibility of $T_{RH}>\MR>\MBL$ which has been known to produce undetectable gravitational signals. For this, we refer the readers to the recent study in \cite{Dunsky:2021tih} .

\begin{figure}[htp]
\centering
\includegraphics[width=11cm]{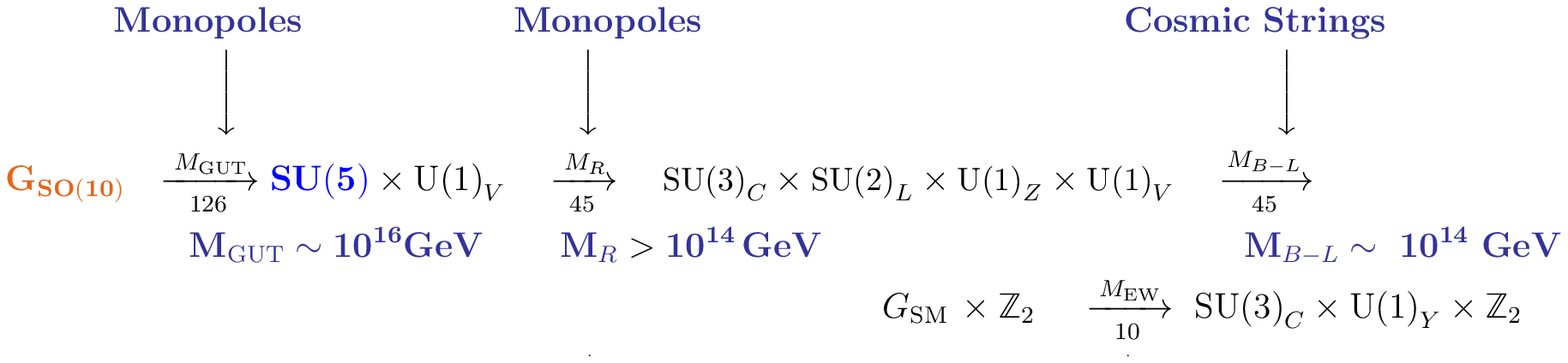}\\
\vspace*{0.3cm}
\includegraphics[height= 7cm,  width=10cm]{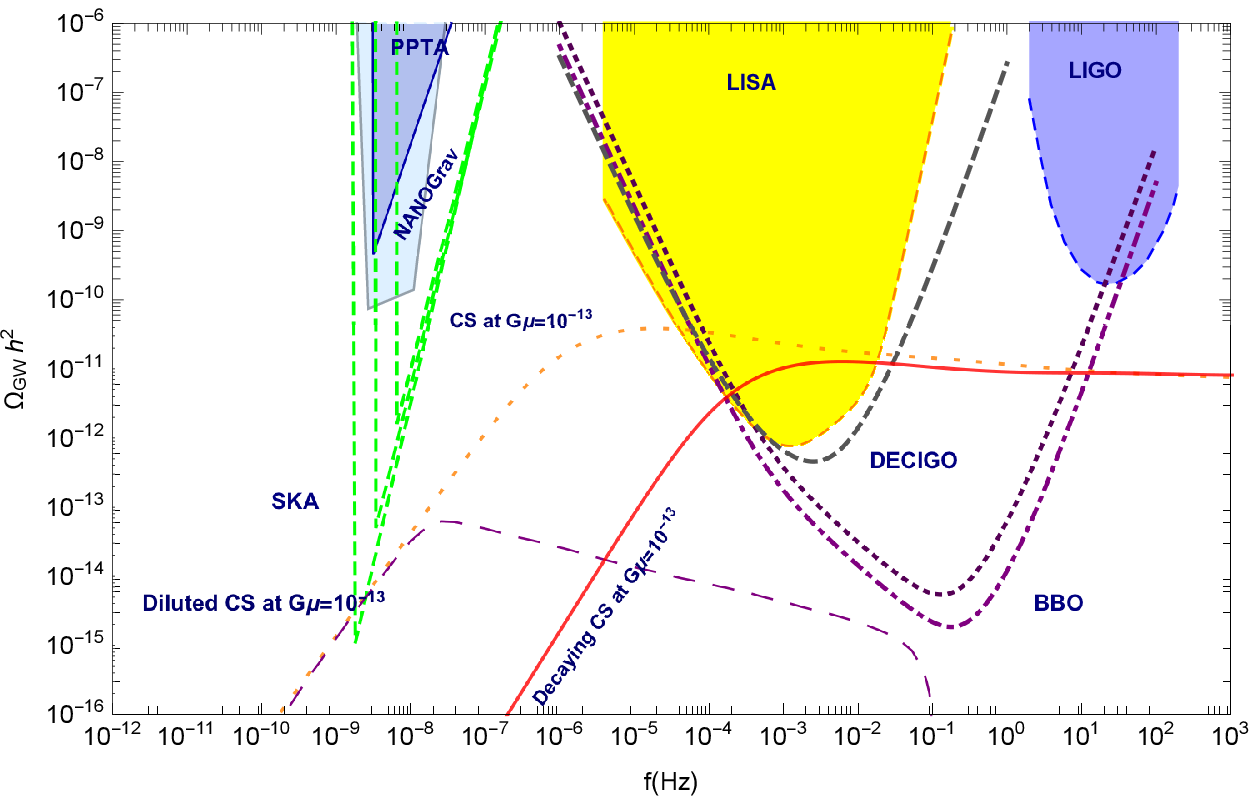}
\caption{The signals from the breaking of supersymmetric SO(10) via an SU(5) route depending on when the inflation takes place.  Taking $\MBL=10^{14}$ GeV we show the usual stable CS signal by the dotted orange line, and   the diluted CS signal by the dashed purple line.  The solid red line shows the decaying CS signal with $\MBL= 2.5 \times 10^{14}$ GeV and  $\sqrt{k}=7.5$ (see \eq{DR:CS}).  For information about the experimental backgrounds please check Appendix \ref{sec:conf_to_GW_Exp}. \label{fig:su5EA}}
\end{figure}

\subsection{$\mathbf{SU(3)}_C \times \mathbf{SU(2)}_L \times \mathbf{SU(2)}_R $ $\times \mathbf{U(1)}_{B-L} $\label{sec:2L2R3BL}}

\paragraph{Phenomenological Considerations}
Given the breaking chain and the Higgs representations, we use the beta functions given in Appendix \ref{app:BetaFunct} to determine the values of $\MBL\!=\!\MR$ and  $\MG$:
\bea
\label{eq:210_Scales}
\MR^{2\, \rm{loop}} = (4.2\pm 0.3) \times 10^{13}\, \rm{GeV},\nonumber\\
\MG^{2\, \rm{loop}} = (1.4\pm 0.26) \times 10^{16}\, \rm{GeV},
\eea
where we have computed the uncertainties based only on the parameters of \Tabref{tbl:tablMzvals}. 
Adapting the prescription from \cite{Fukuyama:2004xs} to our model, we find that the proton decay rate is 
controlled by the superpotential
\bea
W_Y= \left(Y_{10}^{ij} \mathbf{10}_{\bar{T}} + Y_{126}^{ij} \overline{\mathbf{126}}_{\bar{T}} \right)\,  \left(q_i l_j + u_i^c \, d^c_j \right) + 
 \,  
 \left(Y_{10}^{ij} \mathbf{10}_{T} + Y_{126}^{ij} \overline{\mathbf{126}}_{T} \right)
 \left(\frac{1}{2} q_i q_j + u^c_i e^c_j+ d_i^c \nu^c_j  \right),
\eea
where $\mathbf{10}_{\bar{T}}=(\bar{3},1,1,0) + (6,1,1,0)$,  $\mathbf{10}_{T}=(3,1,1,0)+(6,1,1,0)$,  $\mathbf{126}_{\overline{T}}=(\bar{3},1,1,0) + (6,1,1,0)$, and  $\mathbf{126}_{T}=(3,1,1,0)+(6,1,1,0)$, where we have decomposed the SO(10) representations in their $\SUTH_C\times \SUT_L\times \SUT_R\times \UO_{{B-L}}$ decomposition.  All of these representations contain the color triplets and anti-triplets whose masses are denoted by $M_T$.
To keep supersymmetry unbroken at this stage an additional $\mathbf{126}$ is added. 
With these modifications we can write the effective dimension-5 operators \footnote{The dimension-5 effective Lagrangian mediating proton decay is given by 
$\mathcal{L}_5^{\rm{eff}}=C_{L}^{ijkl} O^L_{ijkl} + C_{R}^{ijkl} O^R_{ijkl}$, where $O^L_{ijkl}=\int d^2\theta \epsilon_{abc} (Q^a_i\cdot Q^b_j)(Q^c_k\cdot L_l)/2 $ and
$O^R_{ijkl}=\int d^2\theta \epsilon^{abc}\bar{U}_{ia}\bar{E}_j \bar{U}_{kb}\bar{D}_{lc}$, and the supersymmetric multiplets $Q$, $D$, $E$, $U$ and $L$ are integrated over the Grassmann variable $\theta$.} as in Eq.~(6.13) of \cite{Fukuyama:2004xs}, which is expressed in the following way:
\bea
\label{eq:CLCR126}
C_L^{ijkl} &=& \left(Y_{10}^{ij}, Y_{126}^{ij} \right) M_T^{-1} \left( 
\begin{array}{c}
Y_{10}^{kl},\\
Y_{126}^{kl}
\end{array}
\right),\nonumber\\
C_R^{ijkl} &=& \left(Y_{10}^{ij}- f_1\, Y_{126}^{ij} , f_2\, Y_{126}^{ij} \right) M_T^{-1}  \left( 
\begin{array}{c}
Y_{10}^{kl},\\
Y_{126}^{kl}
\end{array}
\right).
\eea
where $f_1$ and $f_2$ are functions of the vacuum expectation values of $\mathbf{210}$ with $v_R=0$ [see Eq.~(6.8) of \cite{Fukuyama:2004xs}].  The Yukawa couplings $Y_{10}$ and $Y_{126}$ can be written as a linear combination of  the up and down quark Yukawa matrices $Y_u$ and $Y_d$:
\bea
\label{eq:linearcombY}
\left(
\begin{array}{c}
Y_{10}\\
Y_{126}
\end{array}
\right)= M_Y 
\left(
\begin{array}{c}
Y_u\\
Y_d
\end{array}
\right),
\eea
where $M_Y$ contains the coefficients expressing $Y_{10}$ and $Y_{126}$ as a linear combination of $Y_u$ and $Y_d$.
Once the eigenvalues of the Triplet-Higgs mass $M_T$ are found, we use \eq{eq:linearcombY} to rewrite the effective dimension-5 proton decay operators as 
\bea
\label{eq:effWC12610}
C_{L}^{ijkl}&=&\sum_{m=1}^4 \left( Y_u^{ij} A_m  Y_{u}^{kl} \frac{1}{M_{T_m}} +  Y_u^{ij} B_m  Y_{d}^{kl} \frac{1}{M_{T_m}}\right), \nonumber \\
C_{R}^{ijkl} &=& \sum_{m=1}^4 \left(Y_u^{ij} C_m  Y_{u}^{kl} \frac{1}{M_{T_m}} + Y_u^{ij} D_m  Y_{d}^{kl} \frac{1}{M_{T_m}} \right),
\eea
where A, B, C, D are functions of the diagonalization matrices involved in \eq{eq:CLCR126} and  \eq{eq:linearcombY}. The input values of the quark Yukawa, we use 
\bea
Y_u={\text{diag}}(m_u,m_c,m_t)/v/\sin\beta, \quad
Y_d=V_{CKM}^*{\text{diag}}(m_d,m_s,m_b)V^\dagger_{CKM}/v/\cos\beta.
\eea
We then employ the well known procedure of running and matching the Wilson Coefficients of the relevant Lagrangian at each step.
The running of the coefficients $C_{5L, 5R}$ from $\MG$ to $\MBL=\MR$ is governed by 
\bea
\label{eq:EvolveWC}
\frac{d \, C_{5 L,\,CR}^{ijkl}(\mu)}{d\, \ln \mu }= \beta^{Yuk}_{L,R}\, C_{5L, \, 5R}^{ijkl}(\mu)\,,
\eea
where $\beta^{Yuk}_{L,R}$ are the beta functions of $\SUTH_C \times \SUT_L \times \SUT_R\times \UO_{{B-L}}$. 
From $\MBL=\MR$ down to $\MEW$,  we have the usual running of the Minimal Supersymmetric Standard Model (MSSM) and  use the procedure employed in \cite{Ellis:2019fwf}. This means that
the coefficients $C_{5R}^{*331i}(\mu)$, $C_{5L}^{jj1k}(\mu)$ are evolved using \eq{eq:EvolveWC} from the GUT scale to the supersymmetry breaking scale $M_{S}$. At $M_S$,  the sfermions are integrated out via the wino- or Higgsino-exchange one-loop diagrams to get the effective Lagrangian
$\mathcal{L}^{\rm{eff.}}_{MSSM}= C_i^{\tilde{H}}\, O_{1i33}+ C_{jk}^{\tilde{W}}\, \widetilde O_{1jj3}$ + $C_{jk}^{\tilde{W}}\, \widetilde O_{j1jk} + \overline{C}_{jk}^{\tilde{W}}\, \widetilde O_{jj1k}$ where the coefficients
$C_i^{\tilde{H}}$ and $C_{jk}^{\tilde{W}}$ are proportional to $C_{5R}^{*331i}(M_{S})$ and $C_{5L}^{jj1k}(M_{S})$, respectively. 
 Finally,  the coefficients of \eq{eq:ApKnuWC} at the hadronic scale, $C_{RL}(usd\nu_\tau)$, $C_{RL}(uds\nu_\tau)$, $C_{LL}(usd\nu_k)$ and $C_{LL}(uds\nu_k)$ are  proportional to $C_2^{\tilde{H}}$, $C_1^{\tilde{H}}$, $C_{jk}^{\tilde{W}}$ and $C_{jk}^{\tilde{W}}$, respectively (as given in Eq.~ 38 of \cite{Ellis:2019fwf}).
In order to make a definite evaluation of the proton decay in the $ K\, \overline{\nu}$ channel we impose  no-scale supergravity conditions at $\MG$.  Then, we look for values of the supersymmetric particles that satisfy the current limits. Taking $M_0=1.6\times 10^4$ GeV, 
$M_{1/2} =2 \times 10^{4}$ GeV and $\tan\beta=3$, we get
\bea
\tau(p \rightarrow K\, \overline{\nu} )= 8.6 \, \times 10^{33}\  \rm{yrs},
\eea
consistent with the values in \Tabref{tbl:tablMzvals}. 

\begin{figure}
\centering
\includegraphics[width=11cm]{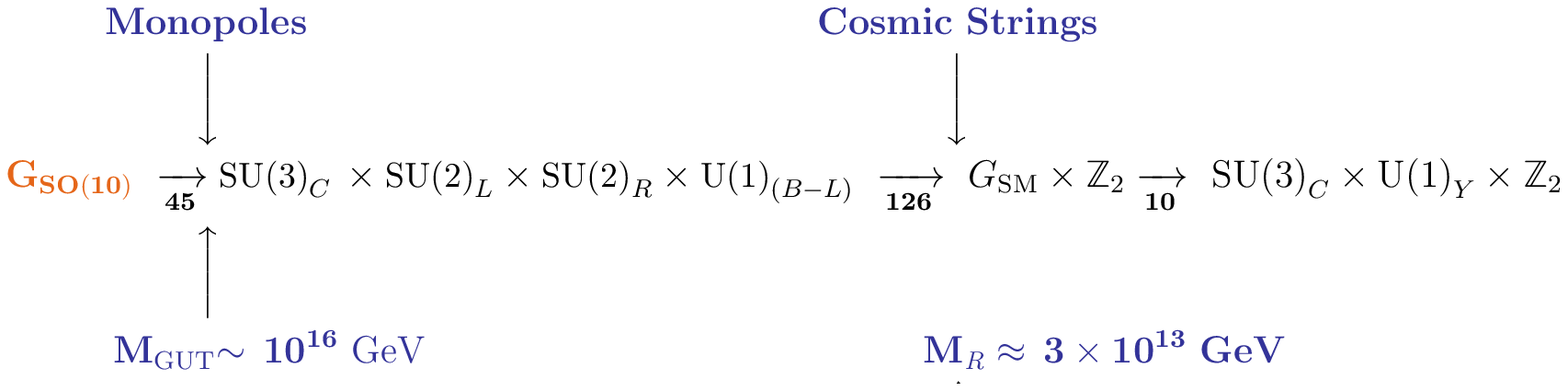}\\
\vspace*{0.5cm}
\includegraphics[height= 5cm,  width=8cm]{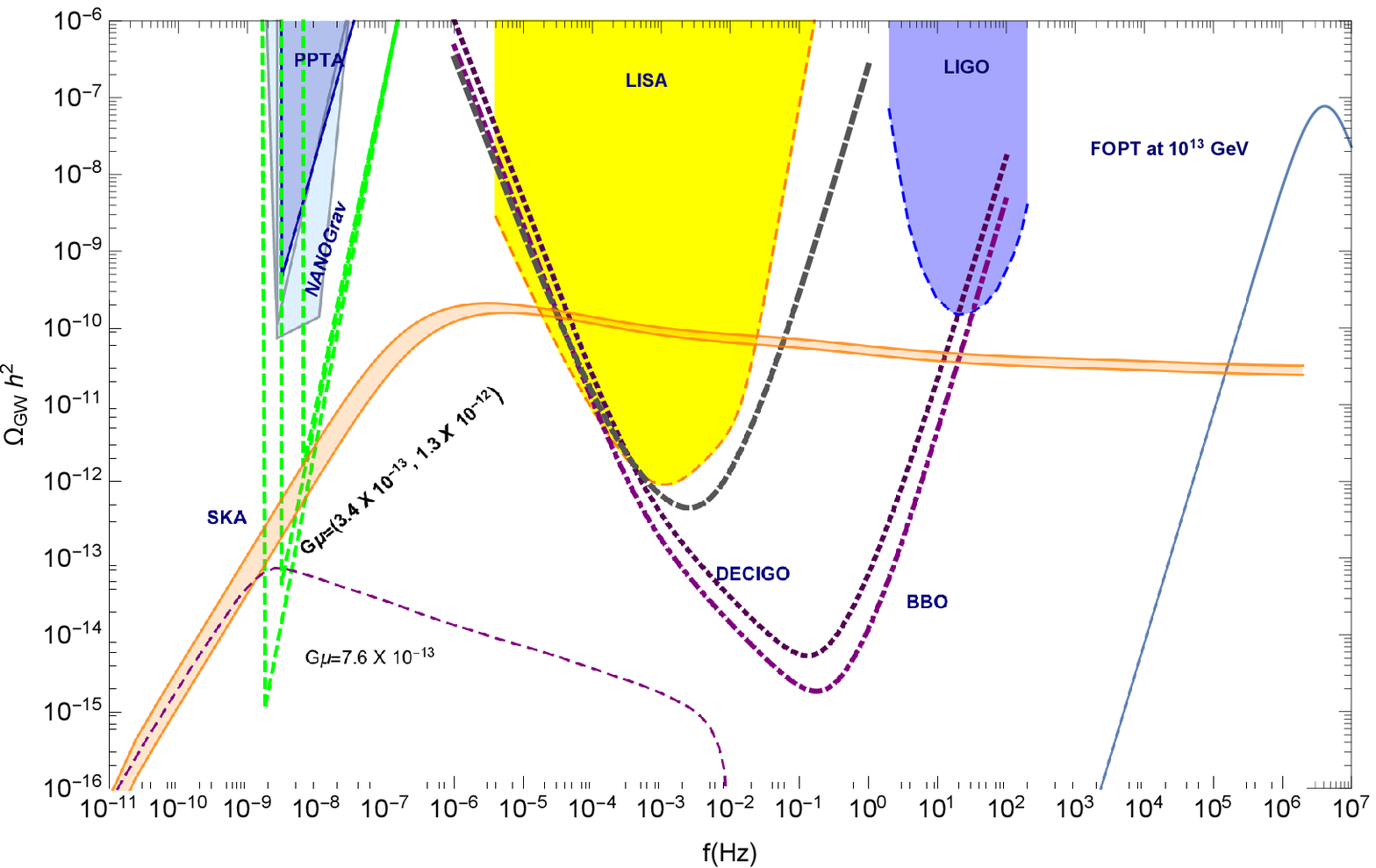}
\includegraphics[height= 5cm,  width=8cm]{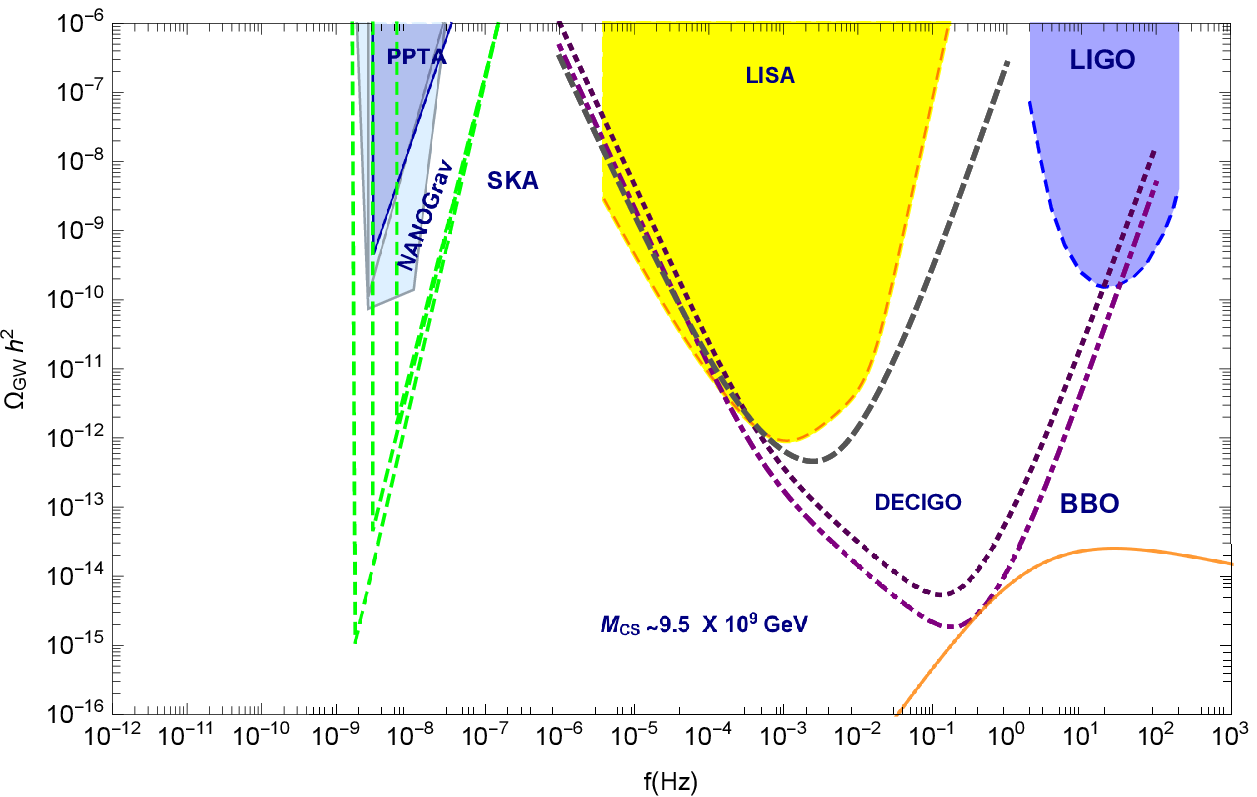}
\caption{ (Left plot: supersymmetric case)
The stable CS signal after inflation ($T_{RH}>\MR$) is shown by 
the orange band corresponding to $\MR =(2,4)\times 10^{13}$ GeV.  The profile of the diluted CS is shown in the dashed purple line with $\MR= 3\times 10^{13}$ GeV (see the text for the details). The decaying CS signal cannot occur due to the large gap between $\MG$ and $\MR$.
The blue solid line shows the signal from FOPT at the scale $10^{13}$ GeV which is, unfortunately, out of the future experimental reach. 
(Right plot: non-supersymmetric case) We show the signals from non-supersymmetric version of the model predicting the intermediate scale at $9.5 \times 10^9$ GeV corresponding to   $G\mu \, =\,  7.6 \times 10^{-20}$. \label{fig:210_susyvsnsusy}}
\end{figure}

\paragraph{First breaking stage} This breaking happens at $\MG$ given in \eq{eq:210_Scales}.  Monopoles are produced at this scale but no CS.

\paragraph{Second breaking stage} At this  scale $\SUT$ and $\UO_{{B-L}}$ are  broken  and  this  can happen at the end of inflation.  
For this case both decaying CS and diluted CS can occur. In the top plot of \Figref{fig:210_susyvsnsusy} we depict 
the breaking chain which predicts metastable strings after inflation at  $10^{13}$ GeV. The scale at which monopoles are produced is $\sim\, 10^{16}$ GeV, and thus the string decay rate is exponentially suppressed by the huge number of $k=v^2/\mu\sim 10^6$ where $v=O(10^{16})$ and $\mu\sim (10^{13} \mbox{GeV})^2$ [see \eq{DR:CS}]. Therefore the strings are almost stable and its signature is plotted in the bottom left panel as an orange band corresponding to $(2,4)\times 10^{13}$ GeV. It may happen also that CS are produced during inflation and thus lead to the GW profile of the diluted CS. This is shown in the left bottom panel as a dashed purple line for a redshift of $\tilde{z}=2\times 10^4$ that satisfies Eq.~(7) of \cite{Cui:2019kkd}.

\paragraph{First Order Phase Transitions} 
Phase transitions have been analyzed in the context of the present breaking chain 
and in the context of no-scale supergravity \cite{Ellis:2020lnc}, and although they are  difficult to realize, there is a possibility and therefore we consider this case.
Coleman-Weinberg inflation inflation could also be adopted to achieve FOPT but this has been explicitly carried out only in the context of non-supersymmetric theories \cite{Lazarides:2021uxv}.

In the bottom-left panel of \Figref{fig:210_susyvsnsusy} we plot a profile of a GW from a FOPT itself at $3\times 10^{13}$ GeV.
Note that the predicted GW profile is way out of the present experiment sensitivities. It will require detectors capable to access a frequency band in the kHz to MHz region. In fact, some proposals of exploring up to kHz are starting to take shape, in particular interferometers \cite{Ackley:2020atn} 
and optically levitated dielectric sensors spanning a wide frequency band from few kHz to $\sim $ 300 kHz \cite{Arvanitaki:2012cn,Aggarwal:2020umq}.  For the MHz region there are also incipient proposals, \cite{Nishizawa:2007tn,Holometer:2016qoh,Martinez:2020cdh} for frequencies up to 100 MHz and in the range 1-250 MHZ \cite{Vermeulen:2020djm}.

The rest of the parameters used to produce the energy density profiles is given in \Tabref{Tbl:FOPTpars} and the relevant formulas are given in Appendix \ref{sec:GWFOPT}.

\begin{table}[h!]
\begin{center}
\begin{tabular}{ |l | l| l |l  |}
\hline
\multicolumn{4}{|c|}{Parameters used for the FOPT profiles}\\
\hline
\hline
 $\alpha$ & $\beta/H$  & $g^*$ & $\epsilon$   \\
\hline
$0.9$  &  1  &   230 & 0.1 \\
\hline
\hline
\multicolumn{4}{|c|}{Parameters used for the DW profile}\\
\hline
\hline
$\mathcal{A}$ & $\tilde{\epsilon}_{GW}$ & $g_{*s}(T_{\rm{Ann.}})$ &  $\sigma$\\
\hline
$0.8$ & $0.3$  & $230$ & $\left( (1-3)\times 10^{14}\, \rm{GeV}\right)^3 $\\
\hline
\end{tabular}
\end{center}
\caption{\small Values of the  parameters used to obtain the FOPT profiles of \Figref{fig:210_susyvsnsusy}
for $\MR= 4.2\times 10^{13}$ GeV, for which we 
choose the temperature equal to the corresponding scale. For the supersymmetric DW profile in \Figref{fig:p126VAR1} we use an annihilation temperature also equal to the a breaking scale of $3\times 10^{14}$ GeV.  
\label{Tbl:FOPTpars}}
\end{table}

\paragraph{Inflation framework}
As mentioned before, this breaking pattern  has  been  analyzed  in \cite{Sarkar:2004ww,Ellis:2016ipm}, and a Starobinsky-like inflation scenario has been constructed successfully with the inflation energy scale $V_{inf}^{1/4} \approx \, 3\times 10^{13}$ GeV which coincides with the scale of the second breaking stage where $B-L$ is broken. Thus, depending on the reheat temperature higher or lower than $\MBL$, one can have undiluted or diluted CS signals.

\paragraph{Comparison to Non-Supersymmetric Case}
  A non-supersymmetric version of this model was considered in \cite{Chakrabortty:2019fov} taking into account threshold corrections. We have checked the values obtained without threshold corrections to confirm
\bea
\label{eq:210_NonSusy_Scales}
\MR^{2\, \rm{loop}} = (9.46\pm 1.0) \times 10^{9}\, \rm{GeV},\nonumber\\
\MG^{2\, \rm{loop}} = (1.60\pm 1.0) \times 10^{16}\, \rm{GeV}.
\eea
It is also shown in \cite{Chakrabortty:2019fov} that the proton decay rate in the channel $p \, \rightarrow \, \pi^0\, e^+$ is above the current bound, $1.6 \times 10^{34}$ yr even without threshold corrections. Future experiments, JUNO\cite{JUNO:2015zny}, DUNE \cite{DUNE:2020ypp} and Hyper-Kamiokande \cite{Hyper-Kamiokande:2018ofw}, have the potential to exclude it. Note however that threshold corrections or additional singlets added to the theory can push-up the limits \cite{Chakrabortty:2019fov}.

In the right plot of \Figref{fig:210_susyvsnsusy} we show the GW signal from a CS (orange curve in the lower right-corner) that would occur at $9.5 \times 10^9$ GeV corresponding to the tension of $G\mu \, =\,  7.6 \times 10^{-20}$). In this non-supersymmetric scenario  inflation can take place before that and so the CS will behave like stable string, which is in the reach of BBO.  

\subsection{$\mathbf{SU(3)}_C \times \mathbf{SU(2)}_L \times \mathbf{SU(2)}_R $ $\times \mathbf{U(1)}_{B-L} \times \mathbf{D}$ \label{sec:su3sec}}

\paragraph{Phenomenological Considerations}
Now we consider the breaking chain depicted in \Figref{fig:p126VAR1}. This allows three  scales of breaking before the breaking to the SM.  The first breaking at $\MG$ is achieved with a $\mathbf{210}$ representation, and the second at $M_D$, where D parity is broken, is achieved with a $\mathbf{45}$ representation. At the last scale,  $\MR$,  both $SU(2)_R$ and  $U(1)_{B-L}$ are broken.  Thus, depending on $T_{RH}$ higher or lower than $M_{B-L}$, one can have undiluted or diluted CS signals.

We recall the reader that there is no unique solution to $\MG, \MD$ and $\MR$. Given one scale, however, the other two are determined. We find specific solutions following the two criteria:
(a) $\MD$ is below $10^{14}$ GeV so that the D parity is broken at a scale compatible with inflation and (b) maximize the GUT scale in order to avoid fast proton decay. 
 The overall ranges of the supersymmetric solutions are
\bea
\label{eq:solrangeEs3}
\MR^{2\, \rm{loop}} &=& (1\times 10^{12},3\times 10^{15})\, \rm{GeV},\nonumber\\
M_D ^{2\, \rm{loop}} &=& (1\times 10^{13}, 1\times 10^{16}) \, \rm{GeV}, \nonumber\\
\MG^{2\, \rm{loop}}& =& (1\times 10^{14}, 2\times 10^{16})\, \rm{GeV}.
\eea
Here, the lowest value of $\MR$ produces the value of $\MG=10^{14}$ GeV leading to too fast proton decay. We note also that the split between $\MR$ and $M_D$ is always $O(10)$, and that once $M_D$ is close to $10^{16}$ GeV it is  indistinguishable from $\MG$.
\footnote{The uncertainty that we quote is for the particular solution, since once $\MR$ is fixed, $M_D$ and $\MG$ are unique and they have to be determined as having convergence of the gauge couplings down to the EW scale within their corresponding uncertainties.}
Out of the possible range of solutions of \eq{eq:solrangeEs3}, we look for a solution where $\MG$ is not smaller than $10^{15}$ GeV to avoid fast proton decay, and $M_R$ is not above $10^{14}$ GeV such that the appearance of DW takes place after inflation.
For these reasons, we focus on the following solution
\bea
\label{eq:selectesolEs3}
\MR^{2\, \rm{loop}} &=& (3.7 \pm 0.2) \times 10^{13}\, \rm{GeV},\nonumber\\
M_D ^{2\, \rm{loop}} &=& (3.8 \pm 0.2) \times 10^{14}\, \rm{GeV},\nonumber\\
\MG^{2\, \rm{loop}}& =& (1.0\pm 0.1) \times 10^{15}\, \rm{GeV},
\eea
where we have included uncertainties from the electroweak parameters, $M_Z$, $\alpha_S$ and $m_t$, quoted in  \Tabref{tbl:tablMzvals}.
We remark that $M_D$ and $\MG$ tend to be closer to each other, than $M_R$ and $M_D$ , mainly because of the value of the coefficient $(b^{(1)}_{2L})_{G_1}$ with respect to $(b^{(1)}_{2L})_{G_2}$, where here $G_2=\SUTH_C\times \SUT_L \times \SUT_R\times \UO_{{B-L}} $ and $G_1=\SUTH_C\times \SUT_L \times \SUT_R\times \UO_{{B-L}}\times D $. These coefficients are such that $(b^{(1)}_{2L})_{G_1} > (b^{(1)}_{2L})_{G_2}$, both for the symmetric and non-supersymmetric cases. The value of $\MG$ for this case, \eq{eq:selectesolEs3}, drives a fast proton decay unless we set the sparticle masses in the 100 TeV range (see \eq{eq:Tp_estimate_gen}). 
At this scale, however the dimension-six operators leading to $p\rightarrow \pi^0 e^+$ become relevant.
To calculate proton decay we proceed as in the previous section \cref{sec:2L2R3BL} with two matching scales, $\MD$ and $\MR$. For the dimension-six operator,  we use the procedure of \cite{Ellis:2019fwf} to  obtain the proton lifetime $2\times 10^{34}$ yrs which is just above the current bound (\Tabref{tbl:tablMzvals}).  For the dimension-five  operator leading to $p\rightarrow K^+ \bar{\nu}$,  we obtain the proton lifetime $9 \times 10^{33}$ yrs with the sparticle masses at 100 TeV, which falls in the reach of the next generation proton decay experiments. 
\begin{figure}
    \centering
    \includegraphics[width=9cm]{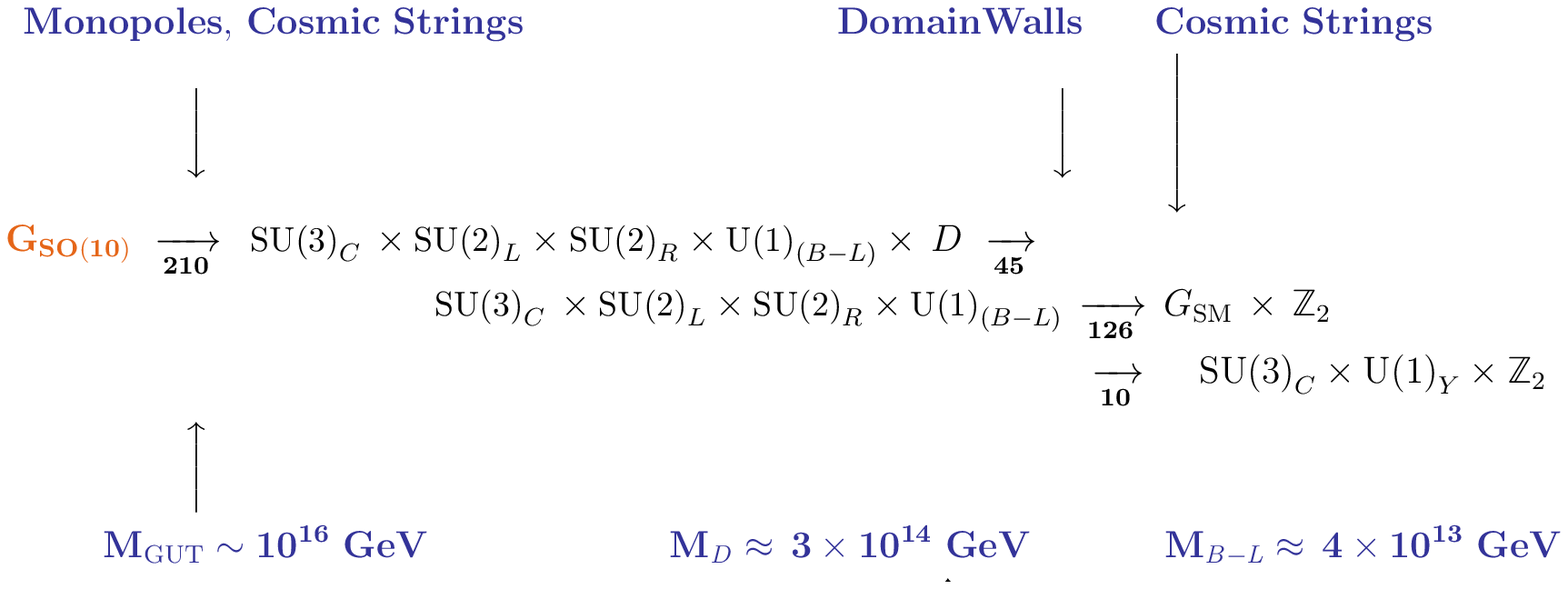}\\
    \vspace*{0.3cm}
    \includegraphics[height= 5cm,  width=8cm]{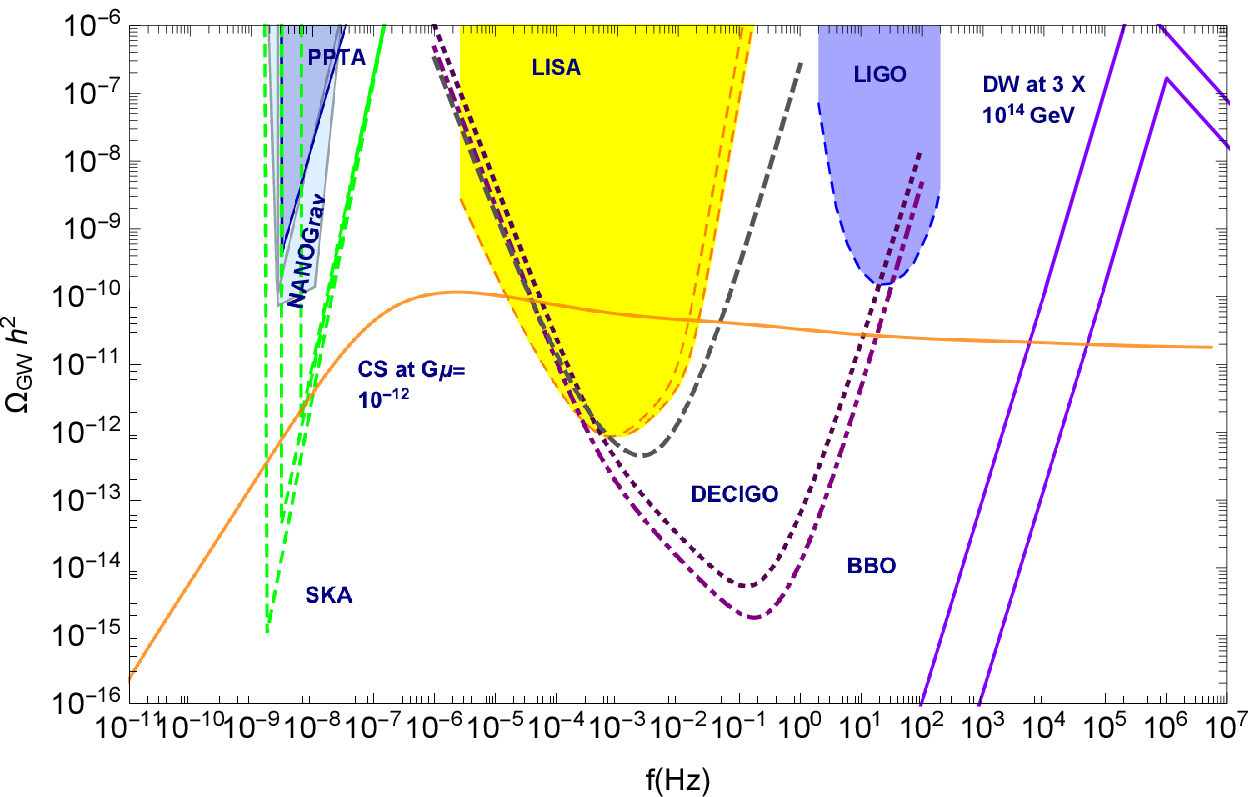}
    \includegraphics[height= 5cm,  width=8cm]{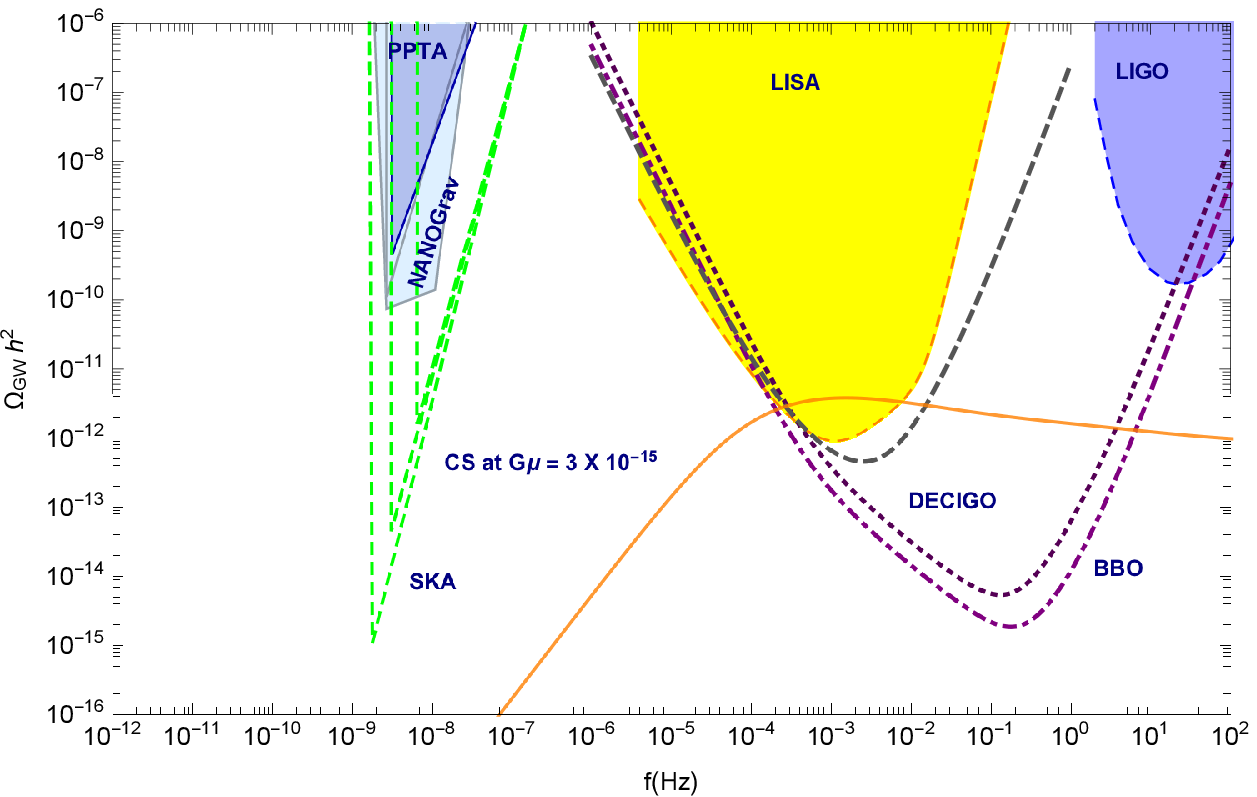}
       \caption{\small (Left plot: supersymmetric case) The solid purple line represents the profile of a GW generated by a DW after inflation at a scale about $10^{14}$ GeV. We have used the parametric formula of \eq{eq:DWparamformula} with the values of \Tabref{Tbl:FOPTpars}. The tension has been taken between $(3 \times 10^{14} \, \rm{GeV})^3$ (top line) and $(1 \times 10^{14}\, \rm{GeV})^3$ (bottom line). The solid orange line is the signal from CS produced at about $3.7 \times 10^{13}\, \rm{GeV}$, corresponding to $G\mu\, =\, 10^{-12}$. 
    (Right plot: non-supersymmetric case) DW are formed above $10^{15}$ GeV and hence are not observable. CS (solid orange line) for this case are produced at $2 \times 10^{12}$ GeV, corresponding to a tension $G\mu$ approximately  equal to $3\times 10^{-15}$.  
    \label{fig:p126VAR1} }
\end{figure}

\paragraph{GW from Domain Walls}
The \emph{BICEP2-Kek Array/Planck} constraint \cite{Akrami:2018odb} on inflation sets an upper bound on the Hubble parameter during single-field inflation at $H_* < 2. 5 \times 10^{-5}\, \MP $ $= 3.0 \times 10^{14} $ GeV. 

This, in turns, sets an upper bound for the scale of this kind of inflation and hence serves as a guide to look for the scales for which topological defects are not completely erased by inflation and reheating.

It is known that the  domain wall network can overclose the universe if it is generated after inflation. This can be overcome in various ways (see Appendix \ref{app:DW}).  One possibility is that the formation of cosmic strings takes place just before inflation and the domain wall formation takes place after inflation. Then the strings can nucleate on the wall, which make domain walls to decay through a quantum tunneling process \cite{Kibble:1982dd,Preskill:1992ck}. This process is controlled by the ratio $\mu/\sigma$ between the string tension $\mu$ and the wall tension $\sigma$. If the radius $R$ of a circular string satisfies $R>\mu/\sigma$, the 
(nucleating a loop of string) rate per unit area is given by \eq{DR:DWnuc}.

In the present case where the D parity breaking produces $\mathbb{Z}_2$ walls, we can use the results of the existing simulations.
These simulations assume  that all damping forces that could be present in the cosmological evolution of the string-wall network are negligible and hence the wall tension goes like $\sigma \sim \MD^3$ \cite{Vilenkin:2000jqa}.

We can then use the formula of  \eq{eq:DWparamformula} to describe the GW signal left by the domain wall.  
Note that in this case an order of magnitude for the hierarchy between $M_D$ and $\MG$  is enough \footnote{In the present case, the domain walls need to be bounded by the strings appearing at the $\MG$ scale and because $\mu[10^{15}]^3> \sigma^2\sim (3.8 \times 10^{14})^2$, then DW profile will resemble that of a stable DW.} to produce a large value of 
$\mu^3/\sigma^2$
which will make the DW evolution indistinguishable from the evolution of DW without the presence of CS.

In this model, the scale $\MD=3.8\times 10^{14}$ GeV 
is of the order of the bound for the Hubble parameter of single-field inflation, as mentioned at the beginning of this section. So in this case the domain walls could have been produced at the end of the inflation leaving their GW as given in \eq{eq:DWparamformula}.
In the left plot of \Figref{fig:p126VAR1} we plot the DW corresponding to this scale with the rest of the parameters as in \Tabref{Tbl:FOPTpars}.

We recall that our motivation is to look for signals of multiple GW sources coming from topological defects or from FOPT from GUTs. So we ask the following question: suppose that a signal corresponding to the shape and frequency of a DW is reconstructed and has a peak frequency of about $10^5$, would this be then indicative of a DW produced by the breaking of a subgroup of a GUT preceded by a breaking producing CS?
To put it in other words, we are asking if such a reconstructed signal would be a clear signal of a DW in the presence of CS.
 To answer the question we recall the following.
 It is known that DW could be also produced when a symmetry is lifted in the vacuum after the breaking of a discrete parity, such that the effective bias term drives the annihilation of the DW \cite{Vilenkin:1981zs,Gelmini:1988sf,Larsson:1996sp}. An specific way to do this is to consider a potential of the form $V=1/4 \lambda \, \phi^4 + \phi^6/\Lambda^2$ \footnote{For a realisation of this scenario in the context of  EW symmetry breaking see for example \cite{Kitajima:2015nla}.}, where  $\Lambda$ is a cutoff scale for the dimension 6 operator and $\phi$ represents the Higgs breaking of the $D$ symmetry. In this case, even in the absence of CS,  such a term could originate from the $\mathbf{16}$ of SO(10), which under the group $\SUTH_C \times \SUT_L \times \SUT_R $ $\times \UO_{{B-L}} \times \mathbf{D}$ contains a singlet and hence both of the terms $\phi^4$ and $\phi^6$ would be allowed.  Then, the tension of the string can be estimated from 
\bea
\sigma \sim V^{1/2}_{\rm{max}}\phi_f,
\eea
where $\phi_f$ represents the value of the Higgs field where the two minima are quasi-degenerated.  At the scale $10^{13}$ GeV the typical values of $V$ and $\phi$ will be respectively $\left(10^{13}\right)^4$ and $10^{13}$, giving then as a result a tension $\sigma$ of the order $\left(10^{13} \, \rm{GeV}\right)^3$. In this way, DW can collapse and part of their energy will be converted to GW. We can take the annihilation temperature to coincide with the scale of the breaking $10^{13}$ GeV, assuming of course that inflation takes place above that scale. The signal will be indistinguishable from the DW signal that we illustrate in \Figref{fig:p126VAR1} we present a DW profile, using the parametric formula of \eq{eq:DWparamformula} with the parameters appearing in \Tabref{Tbl:FOPTpars}.

\paragraph{Comparison to Non-Supersymmetric Case}
Considering the non-supersymmetric model, we find the following ranges for the intermediate scales $\MR$, $\MD$, and $\MG$
\bea
&&\MR^{2\, \rm{loop}} = (1\times10^{10},5\times 10^{13})\, \rm{GeV}, \nonumber\\
&&\MD^{2\, \rm{loop}} = (1\times10^{15}, 1\times 10^{16}) \, \rm{GeV}, \nonumber\\
&&\MG^{2\, \rm{loop}} = (1\times10^{15}, 5\times 10^{16})\, \rm{GeV}.
\eea
These values were obtained following the same procedure as in the supersymmetric case. Notice that we do not find any solution with $\MD$  below $10^{15}$ GeV, and thus $\MD$ is always above the inflation scale.  

The non-supersymmetric counterpart of the model gives the following scales 
\bea
\label{eq:45_Scales}
\MR^{2\, \rm{loop}} &=& (2.0 \pm 0.1) \times 10^{12}\, \rm{GeV},\nonumber\\
\MD^{2\, \rm{loop}} &=& (4.0 \pm 0.1) \times 10^{15}\, \rm{GeV}, \nonumber\\
\MG^{2\, \rm{loop}}& =& (7.3\pm 0.1) \times 10^{15}\, \rm{GeV}.
\eea
which are to be compared with the supersymmetric case \eq{eq:scalessusyEx2}.
 The big differences in the predicted breaking scales will are maintained even after including the full uncertainties. In this nonsupersymmetric case, domain walls would be formed above $10^{15}$ GeV and hence will not be observable. Cosmic strings can be produced after inflation at a scale $2 \times 10^{12}$ GeV, corresponding to approximately to a tension $G\mu= 3\times 10^{-15}$. This is plotted in  \Figref{fig:p126VAR1} with a solid orange line. 
We also checked that the proton lifetime in the $p\rightarrow \pi^0 e^+$ channel is $1.0 \times 10^{35}$ yrs which is above the reach of the next generation of experiments (\Tabref{tbl:tablMzvals}).\\

\section{Conclusions \label{sec:conclusions}}

We studied the possibility of tracking down a route from SO(10) down to the SM in supersymmetric SO(10) models using GW signatures from the topological defects involved in the various stages of breaking, which may differ from the non-supersymmetric counterparts.   Non-supersymmetric models have been studied widely as they require various  intermediate steps of breaking to achieve gauge unification.  Although limited, supersymmetric models may also allow some intermediate breaking scales which could  improve gauge coupling unification. 

The shapes and strengths of gravitational wave signals depend on the scales of intermediate breaking, the scale  of inflation, as well as whether or not hybrid topological defects appear. We explored such features 
considering  the breaking routes of $\SUF$, $\SUTH_C \times \SUT_R \times \SUT_L \times \UO_{{B-L}}$ and $\SUTH_C \times \SUT_R \times \SUT_L \times \UO_{{B-L}} \times D$ with appropriate Higgs representations, and made a comparison between supersymmetric and non-supersymmetric versions.  
For each case, we determined the values of the GUT and intermediate scales by 2-loop RGE, and computed the rate of proton decay in the most constraining channel $p \rightarrow K^+ \bar{\nu}$ assuming no-scale supergravity conditions.
The resulting GW signals depending on the inflation scale are depicted in Figs.~1, 2 and 3.

In the supersymmetric SU(5) example, the only GW signatures that we could identify are from cosmic strings appearing at the intermediate scale around $10^{14}$ GeV.  These could decay due to monopoles predicted in a previous breaking step, or become diluted depending on the scale of inflation. Therefore, variant GW signals are predicted.  The breaking routes of SO(10) via $\SUTH_C \times \SUT_R \times \SUT_L \times \UO_{{B-L}}$ and $\SUTH_C \times \SUT_R \times \SUT_L \times \UO_{{B-L}} \times D$ offer a broader spectrum of GW signatures.

In the case of $\SUTH_C \times \SUT_R \times \SUT_L \times \UO_{{B-L}}$ we chose only one intermediate scale, which turns out to be ${\cal O}(10^{13})$ GeV for the supersymmetric case and ${\cal O}(10^{10})$ GeV for the non-supersymmetric case. As these are much lower than the previous breaking scale at which monopoles appear, the monopoles cannot mediate the decay of cosmic strings. However, diluted cosmic strings could appear depending on the inflation scale. 
 For the  $\SUTH_C \times \SUT_R \times \SUT_L \times \UO_{{B-L}} \times D $ example we chose two intermediate scales: $\MD$ at which D is broken, and  $\MR=\MBL$ below which the SM gauge group is obtained. Unlike in the cases with one intermediate scale, a wide range for the first breaking scale is obtained. We focused on the solution where domain walls appear from the $D$ breaking after inflation, which can occur favorably for the supersymmetric case allowing  $\MD={\cal O}(10^{14})$ GeV, and  $\MR={\cal O}(10^{13})$ GeV. On the other hand,  the non-supersymmetric model predicts  $\MD \gtrsim 10^{15}$ GeV and $\MR \lesssim 10^{13}$ GeV. In both cases stable cosmic strings appear below the scale $\MR$. In the third case, we focused on studying cases where GW from different sources could appear. However, from the range of solutions of \eq{eq:solrangeEs3} it would be possible to choose a value of $\MR=O(10^{14})$ GeV that could correspond to the range at which the IPTA signal has been observed, $G\mu\in$[$2\times 10^{-11}, 3\times 10^{-10}$], but inflation and reheating in this case would need to have finished by then.
 
We conclude by commenting on the reasons why it is difficult to find examples  where GW signatures from different sources can appear. Considering only topological defects, we mainly  need  the breaking of the  GUT group  down to the SM in at least three steps. The separation of these steps needs to be controlled to assure that the topological defects producing observable GW signals occur after inflation and reheating. These features are illustrated in our breaking chain  \ref{sec:su3sec}.  In this case, the appearance of cosmic strings and DW is separated by roughly an order of magnitude, and hence DW decay via nucleation of strings on the wall while avoiding the overclosure of the universe and producing GW at the same time.
Thus, two separate GW signals arise: one from DW at a higher scale and the other from CS at a lower scale.

On the other hand, FOPT could occur at any scale and thus two GW signals can arise at a scale below inflation as shown in \S\ref{sec:2L2R3BL}.
Since monopoles are typically produced at the GUT scale, the scale where GW from FOPT and another source could appear has to be different from the GUT scale and so effectively at least two scales are needed.

\section*{Acknowledgements}
 We thank F. Borzumati, L.  Boubekeur, K. Kaneta, J. Kersten, N. Nagata and K. Olive,  for overall helpful discussions. We also thank K. Kadota and K. Saikawa for clarifications regarding Domain Walls, Q. Shafi for clarifications regarding his works on GW and GUT theories, K. Babu for clarifications about $B-L$ violating proton decay modes;  and specially G. Nardini for very stimulating discussions and information about the LISA agenda.  L. V.-S. acknowledges the support from the ``Fundamental Research Program''  of the Korea Institute for Advanced Study and  the warm hospitality and stimulating environment in these challenging times. The research of L. V.-S. was supported by the National Research Foundation of Korea (NRF) funded by the Ministry of Education through the Center
for Quantum Space Time (CQUeST) with Grant No. 2020R1A6A1A03047877.

\appendix

\section{Confrontation with GW experiments \label{sec:conf_to_GW_Exp}}

The GW signals from stochastic backgrounds are by convention expressed in terms of a GW energy density spectrum $\Omega_{\rm{signal}}(f)\, h^2$ as a function of the GW frequency $f$, while the instantaneous  sensitivity of a GW experiment is expressed as a noise spectrum $\Omega_{\rm{noise}}(f) \, h^2$. 
Then to evaluate if a predicted signal would be able to be detected one can use one of the following procedures.
I. Compute the associated signal-to-noise ratio (SNR) by integrating over the experiments' total observation time accessible to a given frequency \cite{Allen:1996vm,Allen:1997ad,Maggiore:1999vm}: 
 $\text{SNR}^2 = \, n_{\rm{det}}\, t_{\rm{obs}}\, \int^{f_{\rm max}}_{f_{\rm min}}\, df\, \left[\frac{ \Omega_{\rm{signal}}(f)}{\Omega_{\rm{noise}}(f)}\right]^2$,
where $n_{\rm{det}}$, can be 1 or 2 respectively for a cross-correlation or measurement.   Then if SNR is bigger than a threshold value, it is assumed that the associated GW experiment will be able to detect the predicted GW signal.  II. Construct the power-law-integrated sensitivity curve (PLISC), $\Omega_{\rm{PLISC}}$, \cite{Thrane:2013oya} based on $\Omega_{\rm{signal}}(f)$. If the signal and the PLISC intersect in such a way that  $\Omega_{\rm{signal}}(f) > $  $\Omega_{\rm{PLISC}}$ for a given frequency, then is assumed that the experiment will be able to detect the signal. 
However, since the PLISCs  are constructed on spectra based on pure power laws, in realistic situations where the signal is expected to have a structure different to a pure power law, PLISCs it can be used only as a qualitative visual aid. In cases where the shape of the predicted GW signal is fairly model independent the computation of SNR can be computed only once. This fact was exploited in \cite{Schmitz:2020syl}, where new sensitivity curves for SGWB predicted by FOPT were proposed. The shape of these curves  is fairly model independent (the computation depends on the contributions from sound wave and turbulence and their associated parameters). The idea is that using a fit function for a peak-integrated sensitivity curves (PISC) for a particular experiment  it is not  necessary to perform a frequency integration on a parameter-point-by point basis, but simply plot a numerical fit for the PISC against the predicted SWGB. 
Updating only when the spectral shape functions, the noise spectrum and the functional forms of the  peak amplitude change.  
Unfortunately PISC have been obtained only for few experiments in particular LISA, DECIGO and BBO and for SGWB FOPT. We compare our examples in \cref{sec:bcPS} to the  profiles of these last experiments based on the information of \cite{Schmitz:2020syl} and for NANOGrav, PPTA and SKA  we use smoothed profiles for the sensitivity curves based respectively on  \cite{NANOGRAV:2018hou,Arzoumanian:2020vkk}, \cite{Janssen:2014dka} and \cite{Hobbs:2013aka}, while for LIGO whe use the  (PLISCs) \cite{Thrane:2013oya} approach. As a comparison in our plots we plot the LISA  PLISCs and PISC approaches. In all of our plots for LISA, BBO and DECIGO we use $t_{\rm{obs}}=1$ year. For SKA, we present observation times corresponding to one, two and five years, these are shown in the plots appearing in \Figref{fig:su5EA}-\Figref{fig:p126VAR1} respectively as the small, medium and big triangular regions delimited by green dashed lines.

\section{Beta Functions \label{app:BetaFunct}}

\paragraph{$\mathbf{SU(3)_C} \times \mathbf{SU(2)_L} \times \mathbf{SU(2)_R}\times \mathbf{U(1)_{B-L}}$}
For this model the $\mathbf{45}$ breaks the  SO(10) group, breaking D parity as well. The $\overline{\mathbf{126}}$ break subsequently to $G_{SM}\times \mathbb{Z}_2$ and then the $\mathbf{10}$ makes the breaking at the EW scale. Following the order of the indices as $a=2L, 2R, BL, SU(3)_C$, for this model the component acquiring vev in the $\mathbf{126}$ is $(1,3,2,1)$ and that of $\mathbf{10}$ is $(2,2,0,1)$ $\supset (2,1,\pm 1/2)$, this last decomposition being the decomposition under the SM group, $SU(2)_L\times SU(3)_C \times U(1)_Y$.
For the matter, all fermions and their corresponding sparticles in the $\mathbf{16}$ are taken into account: $(2,1,-1/3,3)$, $(2,1,1,1)$, $(1,2,1/3,\overline{3})$ and $(1,2,-1,1)$. Then the beta functions coefficients at one and two loops are respectively 
\bea
\label{eq:bfuncURUBLNoD}
b_a^{(1)}=\left(
\begin{array}{c}
1\\
3\\
21/2\\
-3\\
\end{array}
\right),\quad
b_{ab}^{(2)}=\left(
\begin{array}{cccc}
34 & 8 & 45 & 9 \\
1  & 14 & 9 & 9 \\
15 & 24 & 49 & 3\\
3  & 24 & 3 & 25
\end{array}
\right),
\eea

while for non-supersymmetric, we have
\bea
\label{eq:bfuncURUBLNoDNS}
b_a^{(1)}=\left(
\begin{array}{c}
-3\\
-7/2\\
11/2\\
-7\\
\end{array}
\right),\quad
b_{ab}^{(2)}=\left(
\begin{array}{cccc}
8 & 3 & 3/2 & 12 \\
3  & 80/3 & 27/2 & 12 \\
9/2 & 81/2 & 61/2 & 4\\
9/2  & 9/2 & 1/2 & -20
\end{array}
\right).
\eea
For the non-supersymmetric case our results agree with those of \cite{Mambrini:2015vna,Chakrabortty:2019fov}.
The matching conditions at the scale $\MR$ are as those in \eq{eq:matchingtoMSSM}.

\paragraph{$\mathbf{SU(3)_C} \times \mathbf{SU(2)_L} \times \mathbf{SU(2)_R}\times \mathbf{U(1)_{B-L}} \times \mathbf{D}$}
For this case, the breaking from SO(10) is achieved through a $\mathbf{210}$ representation, which preserves the D symmetry. Then, the $\mathbf{45}$ representation is used to break down to the group $\SUTH_C \times \SUT_R \times \SUT_L \times \UO_{{B-L}}$. The $\mathbf{126}$ is used to break to the MSSM or the SM group. 
Finally the $\mathbf{10}$ representation is used for the final breaking at the EW scale. Specifically, the ${\mathbf{126}}\, \supset \, (1,3,2,1)$ makes the subsequent breaking to the MSSM and the $\mathbf{10}\, \supset \, (2,2,0,1)$ to the SM. The order of the indices is also $a=2L, 2R, BL, \SUTH_C$. The fermions are in the $\mathbf{16}$ representation $ \supset \, (2,1,-1/3,3)$,  $(2,1,1,1)$, $(1,2,1/3,\bar{3})$, $(1,2,-1,1)$. The coefficients of the beta functions are as follows:
\bea
\label{eq:bfuncURUBL}
b_a^{(1)}=\left(
\begin{array}{c}
3\\
3\\
15\\
-3\\
\end{array}
\right),\quad
b_{ab}^{(2)}=\left(
\begin{array}{cccc}
45 & 45 & 61 & 8 \\
3  & 3 & 15 & 24 \\
49 & 49 & 15 & 24\\
9  & 9 & 1 & 14
\end{array}
\right).
\eea
The non-supersymmetric beta function coefficients of this example are
\bea
\label{eq:bfuncURUBL_NS}
b_a^{(1)}=\left(
\begin{array}{c}
-7/3\\
-7/3\\
7\\
-7\\
\end{array}
\right),\quad
b_{ab}^{(2)}=\left(
\begin{array}{cccc}
80/3 & 3 & 27/2 & 12 \\
3  & 80/3 & 27/2 & 12 \\
81/2 & 81/2 & 115/2 & 4\\
9/2  & 9/2 & 1/2 & -26
\end{array}
\right).
\eea
Again for these cases, for the non-supersymmetric case, we reproduce the results of \cite{Mambrini:2015vna,Chakrabortty:2019fov}.
At the scale $\MD$ then the 45 is used to break D parity and from this scale to the scale $\MR=\MBL$ the beta functions of \eq{eq:bfuncURUBLNoD} and \eq{eq:bfuncURUBLNoDNS} are used respectively for the supersymmetric and non-supersymmetric cases.
For both supersymmetric and non-supersymmetric cases, 
the matching from $LRD=\SUTH_C \times \SUT_R \times \SUT_L \times \UO_{{B-L}} \times \mathbf{D}$ to $LR=\SUTH_C \times \SUT_R \times \SUT_L \times \UO_{{B-L}}$ at $\MD$ is simply achieved by 
\bea
g_i^{LRD} = g_i^{LR}.
\eea
The matching of the $LR=\SUTH_C \times \SUT_R \times \SUT_L \times \UO_{{B-L}} \times \mathbf{D}$ gauge couplings to the MSSM/SM gauge couplings  is instead achieved through 
\bea
\label{eq:matchingtoMSSM}
g^{LR}_{2L}(M_I) &=& g^{MSSM/SM}_{2}(M_I), \nonumber\\
g^{ LR}_{B-L}(M_I) &=& z \sqrt{\left(\frac{2}{5\, z^2}  + \frac{3}{5}   \right)}\, g^{MSSM/SM}_1(M_I),\nonumber\\
g^{LR}_{C}(M_I) &=& g^{MSSM/SM}_{3}(M_I),
\eea
here
\bea
g^{LR}_{B-L}(M_I) = z \, g^{LR}_{2R}(M_I),
\eea
where $z$ is a real number that is determined through the constraint of unification at $\MG$.

\paragraph{$\mathbf{SU(4)_C}\times \mathbf{SU(2)_L}\times \mathbf{SU(2)_R}$}

One may consider the breaking route of \eq{bk:case210URBL} and \eq{bk:case45URBL} preceded by the Pati-Salam (PS) group $\SUFR_C\times \SUT_L \times \SUT_R$.
For the case of \eq{bk:case210URBL}, first a supersymmetric version of this chain with the PS group between the
SO(10) group and the $\SUTH_C \times \SUT_L\times \SUT_R$ is not possible to realise with $\MG$ around $(10^{15},10^{16})$ GeV and a minimum matter content. This matter content is with the scalars $10 \supset(1,1,2)$ and $126\supset (1,3,10)$ and the fermions $16\supset (4,2,1) + (\overline{4},2,1)$, where we have given first the matter content in terms of the SO(10) representations and then the PS groups $\SUFR_C\times \SUT_L \times \SUT_R$.
With this content, the one loop beta function coefficients are
$b_{\SUT_L}=1$, $b_{\SUT_R}=21$ and $b_{\SUFR_C}=3$. If the gauge couplings unify, starting from $\MG$, they will all run up in a plane $1/\alpha_i$ vs $\log(\mu/\MG)$, as the energy scale $\mu$ decreases (this happens because all of the values of the beta function coefficients are positive). There is no point where the gauge couplings of
the group $\SUTH_C\times \SUT_L\times \SUT_R \times \UO_{{B-L}}$ can successfully unify, unless the breaking scale of the PS coincides with the $\MG$ scale. 

Non-supersymmetric versions of this model, on the other hand have a wide range of solutions (see Fig. 2, example II4 of \cite{King:2021gmj}) and this owns to the fact that the  coefficients of the one loop beta functions have different signs: $b_{\SUT_L}=-3$, $b_{\SUT_R} = 11/3$ and $b_{\SUFR_C}=-23/3$. This fact then mixes positive and negative slopes in the plane $1/\alpha_i$ vs $\log(\mu/\MG)$ and hence provides a wide range of solutions for $\MG\sim~(10^{14}, 10^{16})$ GeV, $\MPS\approx (10^{13},10^{16})$ GeV and $\MR \sim (10^9, 10^{14})$ GeV.

The case of \eq{bk:case45URBL}, is very similar to the case above. For the supersymmetric version of this chain with the PS group between the SO(10) group and the $\SUTH_C\times \SUT_L\times \SUT_R$ there is no attainable solution with $\MG$ around $(10^{15},10^{16})$ GeV unless $\MG$ and $\MPS$ become the same.
In this case, the minimum matter content is as in the previous case but instead of $126\supset (1,3,10)$, we have $126\supset (1,3,10)+ (3,1,\overline{10})$. The one loop beta function coefficients for this case are $b_{\SUT_L}= b_{\SUT_R}=21$ and $b_{\SUFR_C}=12$, which again are all positive. Non-supersymmetric versions of this chain have on the other hand solutions for $\MG$ close to $10^{16}$ GeV and (see Fig. 2, example III4 of \cite{King:2021gmj}) 
for $\MPS \sim (10^{13}, 10^{15})$ GeV, 
$M_{D} \sim (10^{10},10^{15})$ GeV and
$M_{{B-L}} \sim (10^{11},10^{13})$ GeV.

All of the discussion above could change if of course we consider other than minimal models and add matter content at any step that can affect the value of the beta function coefficients without changing the scalars that define the different breaking patterns.

\section{Production of GW from First Order Phase Transitions \label{sec:GWFOPT}}

Cosmological FOPT originate from a discontinuity in the entropy when there exist a metastable vacuum that eventually decays into the true vacuum of the theory. This event occurs through \emph{bubbling}, that is, the process where regions of space get first to the true vacuum state and overcome the barrier separating the two vacua of the theory. Bubble dynamics produces GW through two basic mechanisms: (i) bubble collisions, generating sound waves, and (ii) turbulence, both producing energy that releases into the GW. In the case of turbulence, it happens when bubble expansion causes macroscopic motion in the surrounding plasma. The total stochastic GW background measured from FOPT is the sum of the contributions from bubble collisions and turbulent motions. However, not all the SGWB are detectable, weak production proceeds via vacuum tunnelling and thermal fluctuations, while strong production happens when bubbles are merely nucleated via quantum tunnelling. Only this latter case is detectable.

Given the breaking of the group into another, producing hence the vacua of the effective theory, the  effective potential relevant for a phase transition is
$\label{eq:effectivepot}
V_{\rm{eff}}(\phi_i,T)= V_0 (\phi_i) + V_{\rm{CW}}(\phi_i) + V^T_1(\phi_i, T)$,
where the fields $\phi$ are all the fields necessary to parameterize the breaking phase, $V_0 (\phi_i)$  is the tree level potential,  $V_{\rm{CW}}(\phi_i)$ is the Coleman-Weinberg contribution (one-loop) and $V^T_1(\phi_i, T)$ is the finite temperature correction.

The parameters carachterising the FOPT are as follows. The ``vacuum-to-thermal energy ratio'', $\alpha$, which is roughly the ratio of the  false vacuum energy density  to the thermal  energy  density, measuring hence the amount of energy released during a FOPT (only for strong production of SGWB $\alpha\ \sim \ O(1)$) 
which is defined as~\cite{Caprini:2015zlo}
$
\alpha=\frac{\Delta\rho}{\rho_R}\;$,
where  $\Delta \rho$ is the released latent heat from the phase transition to the energy density of the plasma background.  The ``change in nucleation rate'',  $\beta$,  is a measure of  the bubble nucleation rate per unit volume, its inverse is approximately the decay rate of the nucleation from the metastable vacuum to the true vacuum and hence it characterises the duration of the FOPT. The other parameters characterising the SGWB  are 
the ``nucleation temperature'', $T_n$ (or  $T_*$), and the velocity of the bubbles, $v_b$. The parameter $\beta$ is given by
\bea
\frac{\beta}{H_n}=T\, \frac{d (S_3(T)/T)}{d T}\, |_{T=T_n}\; ,
\eea
where $S_3$ is the Euclidean action for the $O(3)$-invariant bounce solution.
In terms of the effective potential, 
 the $\alpha$ parameter can be written as
\bea
\alpha (T_n)= \frac{30}{\pi^2 g_* T_n^4} \left(\Delta V_{\eff}(T_n)- T_n\,  \left.\frac{d\, \Delta V_{\eff}(T)}{dT}\right|_{T=T_n}  \right),
\eea
where $\Delta V_{\eff}(T_n)$ is the potential energy difference between the true and the false vacuum at $T_n$.
As mentioned above, the gravitational waves from the FOPT mainly include two sources\footnote{Assuming that the bubbles expand in the plasma reach a relativistic thermal velocity, and hence the contribution due to the scalar field is negligible.}: the bubble collisions, producing sound waves, and the MHD (Magnetohydrodynamics) turbulence, with the total energy given by~\cite{Caprini:2015zlo}
\bea
h^2 \Omega_{\rm{GW}}(f, k_{SW}, k_{\rm{Turb.}})   \approx h^2 \Omega_{\rm SW}  (f, k_{\rm{SW}})\, + \, 
h^2 \Omega_{\rm Turb.}  (f, k_{\rm{Turb.}}).
\eea
The efficiency factors, $k_{\rm{SW}}$ and $k_{\rm{Turb.}}$ characterise the fractions of the released vacuum energy that are converted into the energy of scalar-field gradients, for sound waves and turbulence, respectively. The bubble wall velocity $v_b$ is a function of $\alpha$~\cite{Steinhardt:1981ct}
$
v_b=\frac{1/ \sqrt{3}+\sqrt{\alpha ^2+2 \alpha /3}}{1+\alpha }\;,
$
although this should be taken just as a lower bound  since in phase transitions there exist a larger class of detonation solutions \cite{Laine:1993ey}.
 The energy density of the sound waves that are created in the plasma is given by 
\bea
\Omega h^2_{\rsw}(f)=2.65 \times 10^{-6}(H_*\tau_{\rsw})\left(\frac{H}{\beta}\right)\, v_b
\left(\frac{\kappa_\nu \alpha }{1+\alpha }\right)^2
\left(\frac{g_*}{100}\right)^{-\frac{1}{3}}
\left(\frac{f}{f_{\rsw}}\right)^3 \left(\frac{7}{4+3 \left(f/f_{\rsw}\right)^2}\right)^{7/2},
\eea
where the factor $\tau_{sw}$ is given by $\rm{min}\left[\frac{1}{H_*},\frac{R_*}{\bar{U}_f}\right]$ is the time scale of the duration of the phase. It could be either  $1/H_*$ or  ${R_*}/{\bar{U}_f}$, where  $H_*R_*=v_b(8\pi)^{1/3}(\beta/H)^{-1}$  according to ~\cite{Ellis:2020awk}. The root-mean-square (RMS) fluid velocity is given roughly by \cite{Hindmarsh:2017gnf, Caprini:2019egz, Ellis:2019oqb} $
\bar{U}_f^2~\approx~ \frac{3}{4}\, \left(\frac{\kappa_\nu\alpha}{1+\alpha}\right)$, where $\kappa_\nu$ is the fraction of the latent heat transferred into the kinetic energy of the plasma   \cite{Espinosa:2010hh}\footnote{In the small and large limit of $v_b$,  $\kappa_\nu$ can be approximated as $\kappa_\nu \approx \alpha\, (0.73 + 0.083\, \sqrt{\alpha}+ \alpha)^{-1}$, $v_b\sim 1$ and $\kappa_\nu \approx v_b^{6/5}\, 6.9\, \alpha (1.36 - 0.037 \sqrt{\alpha} + \alpha)^{-1} $ for $v_b \lesssim 0.1 $}.
The peak frequency is given by
\bea
f_{\rsw}=1.9 \times 10^{-5}\, \left(\frac{\beta}{H}\right)\, \frac{1}{v_b} \left(\frac{T_*}{100\, \rm{GeV}}\right) \left({\frac{g_*}{100}}\right)^{\frac{1}{6}} {\rm Hz }\;.
\eea
The energy density of the MHD turbulence in the plasma is given by 
\bea
\Omega h^2_{\turb}\, (f)=3.35 \times 10^{-4}\left(\frac{H}{\beta}\right)\,
\left(\frac{\epsilon\, \kappa_\nu \alpha }{1+\alpha }\right)^{\frac{3}{2}}
\left(\frac{g_*}{100}\right)^{-\frac{1}{3}}\, v_b\, 
\frac{\left(f/f_{\turb}\right)^3\left(1+f/f_{\turb}\right)^{-\frac{11}{3}}}{\left[1+8\pi f \frac{a_0}{(a_* H_*)}\right]}\;,
\eea
where 
\bea
 \frac{a_*}{(a_0)}=\left(\frac{g_0}{g_*}\right)^{1/3}\, \frac{T_0}{T_*} \sim 1.25 \times 10^{-7}\,  T_*(\rm{GeV})\, g_*^{1/6},
\eea
and the efficiency factor is  given by $\epsilon \approx 0.1$. The present day Hubble parameter can be expressed as
\bea
H_{*} = \left( 1.65 \times 10^{-5} Hz \right) \left( \frac{T_{*}}{100 \rm{GeV}} \right) \left( \frac{g_{\ast}}{100} \right)^{1/6}.
\eea
Finally, the peak frequency for GW produced by MHD turbulence is given by \cite{Caprini:2009yp}
\bea
f_{\turb}=2.7~  \times 10^{-5}
\left(\frac{\beta}{H}\right)\, \frac{1}{v_b} \left( \frac{T_*}{100\ \rm{GeV}} \right) \left({\frac{g_*}{100}}\right)^{\frac{1}{6}} {\rm Hz }\;.
\eea
In our study we have considered just the contributions from bubble dynamics (leading) and turbulence (sub-leading).

\section{Production of GW from Cosmic Strings \label{app:ProdGWCS}}

\paragraph{Stable Cosmic Strings}
GUT based on simple gauge groups lead to the formation of topologically stable monopoles whose density is about $10^{18}$ times greater than the experimental limit \cite{Akrami:2018odb}, dominating thus the energy density of our universe and closing it. 
While this kind of topological 
cosmological defects demands then a wash down effect such as inflation, not all monopoles need to undergo completely washout \cite{Vilenkin:1981zs}. 

Cosmic strings can also have enormous energy. In the simplest case, which we consider, the canonical type of string (Nambu-Goto), the energy per unit length, $\mu$, and the string tension are equal.  It is expected that for strings produced at a phase transition at $T_c$, $\mu \sim T_c^2$ \cite{Hindmarsh:1994re}, where $\mu$ is the tension of the string  and it characterizes the strength of the gravitational interaction of strings. A grand unified string of length equal to the solar diameter would be as massive as the sun, while such a length of string formed at the electroweak scale would weigh only 10 mg. The gravitational effects of the latter are essentially negligible, though such strings may still be of great interest, because of other types of interactions. The Nambu-Goto cosmic strings are characterized only by the string tension, $\mu$  \cite{Vilenkin:2000jqa}. Using the  Kibble mechanism, the string tension can be estimated to be (with a GUT scale of $10^{16}$ GeV)~\cite{Gouttenoire:2019kij,Gouttenoire:2019rtn}
  \bea
 \mu\approx \frac{10^{-15}}{ G}\left(\frac{T_p}{10^{12}~ {\text {GeV}}}\right)^2\;,\label{eq:roughmu}
 \eea
 where $G$ is the Newton's constant. One simple approximation is to assume $T_p\approx T_n$.  When this scale is taken to be the GUT scale, roughly $T_p=T_n=10^{16}$ GeV, we then get the result of CS produced at GUT scale, we get then the familiar result
\bea
G\, \mu \approx 10^{-7}.
\eea
Another way to express the tension of the CS is to write \cite{Hill:1987qx,Hindmarsh:2011qj}
\bea
\label{eq:MCSIntT}
G \mu = \frac{1}{8} \, B(x) \, \left(\frac{M_{\rm{CS}}}{\MP} \right)^2,
\eea
where $M_{\rm{CS}}$ is the scale of the cosmic string and $\MP$ the reduced Planck scale, $2.4 \times 10^{18} $ GeV. The function $B(x)$ is $B(x)=1.04\, x^{0.195}$ for $10^{-2}\, \ll x \ll 10^2$, $B(x)=2.4\, \log(2/x)$ for $x\lesssim 0.01$.
 After the formation of strings, the string loops loss energy dominantly through emission of gravitational waves.
We compute the relic GW energy density spectrum from cosmic string networks from \cite{Cui:2017ufi,Cui:2018rwi}
\bea
\label{eq:GWd1}
\Omega_{\rm GW}(f) =\sum_k \Omega_{\rm GW}^{(k)}(f),
\eea
where  
\bea
\label{eq:GWd2}
\Omega_{\rm GW}^{(k)}(f) =
\frac{1}{\rho_c}
\frac{2k}{f}\, \frac{\Gamma\, k^{-4/3}}{\zeta(4/3)}
\frac{\mathcal{F}_{\alpha}\, G\mu^2}
{\alpha\left( \alpha+\Gamma G\mu\right)}
\int_{t_F}^{t_0}\!d\tilde{t}\;
\frac{C_{eff}(t_i^{(k)})}{t_i^{(k)\,4}}
\left[\frac{a(\tilde{t})}{a(t_0)}\right]^5
\left[\frac{a(t^{(k)}_i)}{a(\tilde{t})}\right]^3
\,\Theta(t_i^{(k)} - t_F),
\eea
the critical density is given by $\rho_c= 3H_0^2/8/\pi/G$, $k$ is the k-mode at a frequency f.  The gravitational loop-emission efficiency factor is $\Gamma\approx50$ \cite{Blanco-Pillado:2017oxo} with its Fourier modes for cusps \cite{Olmez:2010bi} given by \cite{Blanco-Pillado:2013qja,Blanco-Pillado:2017oxo}
$
\frac{\Gamma\, k^{-4/3}}{\zeta(4/3)}=
\Gamma^{(k)} = \frac{\Gamma k^{-\frac{4}{3}}}{\sum_{m=1}^{\infty} m^{-\frac{4}{3}}}$, 
where $\sum_{m=1}^{\infty} m^{-\frac{4}{3}} \simeq 3.60$.
The factor $\mathcal{F}_{\alpha}$ characterizes the fraction of the energy released by long strings. We use $\mathcal{F}_{\alpha}=0.1$, and 
 $\alpha = 0.1$ in order to take into account the length of the string loops rendering a monochromatic loop distribution. $\Theta$ is the Heavy side step function and $a(t)$ is the cosmological factor at a given time $t$. The loop production efficiency $C_{eff}$ is obtained after solving Velocity-dependent One-Scale equations (VOS),  with $C_{eff}=5.4(0.39)$ in radiation (matter) dominate universe~\cite{Gouttenoire:2019kij}. The VOS equations are
\bea
\label{eq:voseqs}
\frac{dL}{dt} &=&(1+\overline{v}) \, HL + \tilde{c}\, \frac{\bar{v}}{2}\,,\nonumber\\
\frac{d\overline{v}}{dt} &=& \left(1-{\overline{v}}^2\right) \left[ \frac{k(\overline{v})}{L}- 2H \overline{v} \right]\,,
\eea
where $k(\overline{v})= 2\sqrt{2}/\pi \, (1-{\overline{v}}) (1+ 2\sqrt{2} {\overline{v}}^2) \left(\frac{1-8\, {\overline{v}}^6 }{1+8\, {\overline{v}}^6} \right)  $ and $\tilde {c} \approx 0.23$ describes loop formation. $L$ is the correlation length parameter (such that the energy density of the long strings is given by $\rho_\infty=\mu/L^2$ \cite{Vilenkin:2000jqa}) and $H$ is the corresponding Hubble parameter. The scaling regime is reached after three or four orders of magnitude of change in the energy scale of the universe, where we have a stable value of  $C_{eff}$, see e.g. Fig.3 of \cite{Gouttenoire:2019kij}. 
The loop formation time of the k mode is a function of the GW emission time ($\tilde{t}$) and is given by
\bea
t_i^{(k)}(\tilde{t},f) = \frac{1}{\alpha+\Gamma G\mu}\left[
\frac{2 k}{f}\frac{a(\tilde{t})}{a(t_0)} + \Gamma G\mu\;\tilde{t}
\right].
\eea
Assuming the small-scale structure of loops is dominated by cusps, the high mode in \eq{eq:GWd1} is given by
$\Omega_{\rm GW}^{(k)}(f)
= \frac{\Gamma^{(k)}}{\Gamma^{(1)}}\,\Omega_{\rm GW}^{(1)}(f/k)
=k^{-4/3}\,\Omega_{\rm GW}^{(1)}(f/k)$.
The low and high frequencies of the spectrum of the GW from cosmic strings are dominated by emissions in the  matter and radiation dominated eras respectively.

\paragraph{Diluted Cosmic Strings }
A standard expectation of primordial cosmological inflation is that it dilutes all relics created to unobservable levels. But this does not need to be the case, for example in \cite{Cui:2019kkd,Guedes:2018afo} counter-examples were presented. These correspond to  networks of cosmic strings diluted by inflation but that can regrow to a level potentially observable today in gravitational waves (GWs).
 In contrast to undiluted cosmic strings (from a stochastic GW background),  the leading signal from a diluted cosmic string network can be distinctive bursts of GW. In \cite{Cui:2019kkd} the VOS model was used together with a simplified picture of inflation and reheating to estimate the dilution of cosmic strings. 
The starting point is to consider the same VOS equations as in \eq{eq:voseqs}. Then  take the correlation length $L_F(t_F)$ at the time $t_F$, the greater of the beginning of inflation or the formation of the network. After $t_F$, the string network parameters reach an attractor solution during inflation given by $L(t)=L_F e^{H_I(t-t_F)}$ and $\overline{v}(t)=2\sqrt{2}/\pi/H_I/L(t)$ \cite{Cui:2019kkd}. The solution corresponds to having the long string network with with $HL \gg 1$ and $\overline{v}\ll 1$. The conditions under which there is enough regrowth are given in Eq.~(7) of \cite{Cui:2019kkd}. These are written in terms of the redshift, $\tilde{z}$, \cite{Cui:2019kkd} at which the condition $HL\rightarrow 1$ is achieved. The condition can be satisfied for $\Delta N \neq 0$ , where  $\Delta N$ represents the number of e-foldings between $t_F$ and $t_I$, the time at which the attractor solution is satisfied.  $\Delta N=0$ corresponds to the string forming at the start of inflation  and in this case then only the number of long strings accounts for satisfying the condition of Eq.~(7) of \cite{Cui:2019kkd}. We assume this last possibility and use Eq.~(19) of \cite{Cui:2019kkd} to produce the GW profile.

\section{Production of GW from Domain Walls \label{app:DW}}

 Domain walls are sheet-like topological defects, which might be created in the early universe when a discrete symmetry is spontaneously broken. Since discrete symmetries are ubiquitous in high energy physics beyond the Standard Model (SM), many new physics models predict the formation of domain walls in the early universe. By considering their cosmological evolution, it is possible to deduce several constraints on such models even if their energy scales are much higher than those probed in the laboratory experiments.

In the context of GUT theories they can appear via the D-parity symmetry, denoted also as $Z^C_2$. They cannot appear after inflation has taken place, unless they are broken by an explicit parameter, by  gravitational effects, or decay due to the attachment of CS produced in a previous breaking stage. 
For DW to which CS do not attach, there could be also the possibility that the discrete symmetry is broken but then restored at a higher temperature \cite{Weinberg:1974hy}. 
Another possibility is to accompany the GUT theory with an extra PQ symmetry that lifts the symmetry  $Z_2^C$ and so it effectively only acts like an effective symmetry (see for example \cite{Saikawa:2017hiv} and references within it). 

For some ranges of parameters there exists an alternative solution to the domain wall problem \cite{Dvali:1995cc,Dvali:2018txx} based on the idea of symmetry non-restoration \cite{Weinberg:1974hy}, which does not require any explicit breaking of the discrete symmetry.  Simply because the discrete symmetry is never restored at high temperature. In this way the domain walls never form. As it is shown in \cite{Dvali:1995cc}, this can rescue some of the models such as those with spontaneously broken CP and Peccei-Quinn symmetries. Nevertheless, this mechanism is incompatible with renormalizable supersymmetric theories \cite{Mangano:1984dq}. One way to make the bias compatible with quantum breaking \cite{Dvali_2014, Dvali:2014gua} and with the de Sitter Swampland program \cite{Garg:2018reu, Ooguri:2018wrx}, would be to make a bias time-dependent in such a way that it disappears after the walls disappear \cite{Dvali:2018txx}.  Nevertheless, we think that identification or lack of signals of domain walls in the expected regions would lead to either constrain of rule out the parameter space of bias parameters. 

Assuming that a bias parameter, coming  either by Planck suppressed terms, \cite{Rai:1992xw, Mishra:2009mk}  or by a soft breaking, breaks the symmetry and hence allows D parity to break below the inflation scale leaving a GW signal described by
\bea
\label{eq:DWparamformula}
\Omega_{\rm{DW}}\, h^2(f)&=& 
7.2 \times 10^{-18}\, {\mathcal{A}}^2  \, \tilde\epsilon_{GW}
{\left(\frac{g_{*s (T_{\rm{Ann.}})}}{10}\right)^{-4/3} \left( \frac{\sigma}{1\, \rm{TeV}^3} \right)^2 }{
\left(\frac{T_{\rm{Ann.}}}{10^{-2}\, \rm{GeV}}\right)^4}\, \frac{f^p}{f_{\rm{Peak}}^p},\\
& & \hspace*{3cm} f\ \left\{
\begin{array}{l}
\leq f_{\rm{Peak}},\quad p = 3,\nonumber\\
> f_{\rm{Peak}},\quad p = -1
\end{array}
\right.,
\eea
where we have used the parametric form of the frequency below and above the peak and the assumptions of \cite{Saikawa:2017hiv} based on the simulations of \cite{Hiramatsu:2013qaa}. $\mathcal{A}$ is a parameter that can be extracted from lattice simulations, $\tilde{\epsilon}_{\rm{GW}}$ is a parameter based on numerical simulations for the energy density of the GW and it has a value $\tilde{\epsilon}_{\rm{GW}}= 0.7\pm 0.4$  \cite{Saikawa:2017hiv}, ${g_{*s (T_{\rm{Ann.}})}}$ are the degrees of freedom at the annihilation temperature, $T_{\rm{Ann.}}$. Finally $\sigma$ is the tension of the DW. Note that a DW which develops after inflation and during the radiation domination era and where cosmic strings formed before inflation will have the same parametric behaviour as \eq{eq:DWparamformula},  that is 
$
\Omega_{\rm{DW}}\, h^2(f)= \Omega_{\rm{GW}, \, \rm{max.}} \,  \frac{f^p}{f_{\rm{Peak}}^p}$,
where the peak frequency is determined by the Hubble parameter at the decay time, $f_{\rm{Peak}} \sim a_{t_{\rm{dec.}}}/a_{t_0}\, H_{\rm{t_{dec.}}}$ \cite{Hiramatsu:2013qaa}.

\printbibliography



\end{document}